\documentclass[a4paper,article]{cas-dc-1}
\usepackage{amsmath,amsfonts}
\usepackage{array}
\usepackage[caption=false,font=normalsize,labelfont=sf,textfont=sf]{subfig}
\usepackage{textcomp}
\usepackage{stfloats}
\usepackage{url}
\usepackage{verbatim}
\usepackage{graphicx}
\usepackage{nomencl}
\usepackage{etoolbox}
\usepackage{ifthen}
\usepackage{booktabs}
\usepackage[ruled]{algorithm2e}
\usepackage{appendix}
\usepackage{multirow}
\usepackage{bm}
\usepackage{makecell}
\usepackage{algpseudocode}
\allowdisplaybreaks 
\usepackage[subtle,tracking=normal]{savetrees}
\usepackage{float}
\usepackage{caption} 

\usepackage[numbers,sort&compress]{natbib}
\usepackage{nomencl}
\renewcommand{\nomgroup}[1]{%
    \ifthenelse{\equal{#1}{A}}{\item[\textbf{Binary variables}]}{%
    \ifthenelse{\equal{#1}{B}}{\item[\textbf{Continuous variables}]}{%
    \ifthenelse{\equal{#1}{C}}{\item[\textbf{System input data/parameters}]}{%
    \ifthenelse{\equal{#1}{D}}{\item[\textbf{DAC input parameters}]}{%
    \ifthenelse{\equal{#1}{N}}{\item[\textbf{Abbreviations}]}{}}}}}}
\usepackage{multicol}
\usepackage{framed} 
\usepackage{etoolbox} 
\usepackage{float}
\usepackage{placeins}
\usepackage{threeparttable} 
\usepackage{subcaption}

\def\tsc#1{\csdef{#1}{\textsc{\lowercase{#1}}\xspace}}
\tsc{WGM}
\tsc{QE}

\newcommand{\rev}[1]{\textcolor{black}{#1}}
\newcommand{\co}[0]{CO\textsubscript{2} }
\newcommand{\cd}[0]{CO\textsubscript{2}}

\makeatletter
\RequirePackage{expl3}
\ExplSyntaxOn
\cs_if_exist:NF \__first_footerline:  { \cs_new:Npn \__first_footerline:  {} }
\cs_if_exist:NF \__second_footerline: { \cs_new:Npn \__second_footerline: {} }
\cs_if_exist:NF \__first_headerline:  { \cs_new:Npn \__first_headerline:  {} }
\cs_if_exist:NF \__second_headerline: { \cs_new:Npn \__second_headerline: {} }
\ExplSyntaxOff
\makeatother

\begin{document}
\let\WriteBookmarks\relax
\def\floatpagepagefraction{1}
\def\textpagefraction{.001}
\let\printorcid\relax %

\shorttitle{\rmfamily Author’s preprint version, submitted to \textit{Engineering}. Not the final published version.}    

\shortauthors{\rmfamily Z. Fan et al.}

\title[mode = title]{Enhancing Profit and \co Mitigation: Commercial Direct Air Capture Design and Operation with Power Market Volatility}

\author[1,2]{Zhiyuan Fan}
\cormark[1]
\ead{zf2198@columbia.edu} 
\credit{Conceptualization, Modeling, Methodology, Software, Writing}

\author[3]{Elizabeth Dentzer}
\credit{Modeling, Methodology, Software}

\author[4]{James Glynn}
\credit{Supervision, Writing, Revision, Funding Support}

\author[5]{David S. Goldberg}
\credit{Writing, Revision}

\author[6]{Julio Friedmann}
\credit{Writing, Revision}

\author[1]{Bolun Xu}
\credit{Conceptualization, Methodology, Supervision, Revision, Funding Support}


\address[1]{Department of Earth and Environmental Engineering, Columbia University, New York, NY 10027, USA}
\address[2]{Center on Global Energy Policy, Columbia University, 1255 Amsterdam Avenue, New York, NY 10027, USA}
\address[3]{Department of Mechanical and Aerospace Engineering, Princeton University, NJ 084540, USA}
\address[4]{Energy Systems Modelling Analytics Limited, Galway, Ireland}
\address[5]{Lamont-Doherty Earth Observatory, Columbia University, 61 Rte 9W, Palisades, New York, NY 10964, USA}
\address[6]{Carbon Direct, 17 State Street, New York, NY 10004, USA}

\cortext[1]{Corresponding author} 

\begin{abstract}
Current decarbonization efforts are falling short of meeting the net-zero greenhouse gas (GHG) emission target, highlighting the need for substantial carbon dioxide removal methods such as direct air capture (DAC). However, integrating DACs poses challenges due to their enormous power consumption. This study assesses the commercial operation of various DAC technologies that earn revenue using monetized carbon incentives while purchasing electricity from wholesale power markets. We model four commercial DAC technologies and examine their operation in three representative locations including California, Texas, and New York.
Our findings reveal that commercial DAC operations can take financial advantage of the volatile power market to operate only during low-price periods strategically, offering a pathway to facilitate a cost-efficient decarbonization transition. The ambient operational environment such as temperature and relative humidity has non-trivial impact on abatement capacity. Profit-driven decisions introduce climate-economic trade-offs that might decrease the capacity factor of DAC and reduce total \co removal. These implications extend throughout the entire lifecycle of DAC developments and influence power systems and policies related to full-scale DAC implementation. Our study shows that DAC technologies with shorter cycle spans and higher flexibility can better exploit the electricity price volatility, while power markets demonstrate persistent low-price windows that often synergize with low grid emission periods, like during the solar ``duck curve" in California. An optimal incentive design exists for profit-driven operations while carbon-tax policy in electricity pricing is counterproductive for DAC systems.

\end{abstract}



\maketitle

\section{Introduction}

\subsection{Background and motivation}

Current decarbonization efforts are failing short in meeting the net-zero greenhouse gas (GHG) emission target, and studies have shown that the most likely approach will be to overshoot 1.5\textdegree C target followed by more intensive \co removal from the atmosphere~\cite{lee_ipcc_2023,  lamboll_assessing_2023, tollefson2023too, buck_why_2023}. Yet, the scale of potentially required carbon dioxide removal (CDR) deployment is enormous, ranging from 160-370 Gt-\co net removal by 2100 for every 0.1\textdegree C overshoot \cite{lee_ipcc_2023}. Depending on the chosen scenarios, up to 30 Gt-\co of negative emissions per year will be required to balance the global carbon budget \cite{marcucci_road_2017}. 

\begin{figure*}[!ht]
        \centering
        \includegraphics[width=0.55\textwidth]{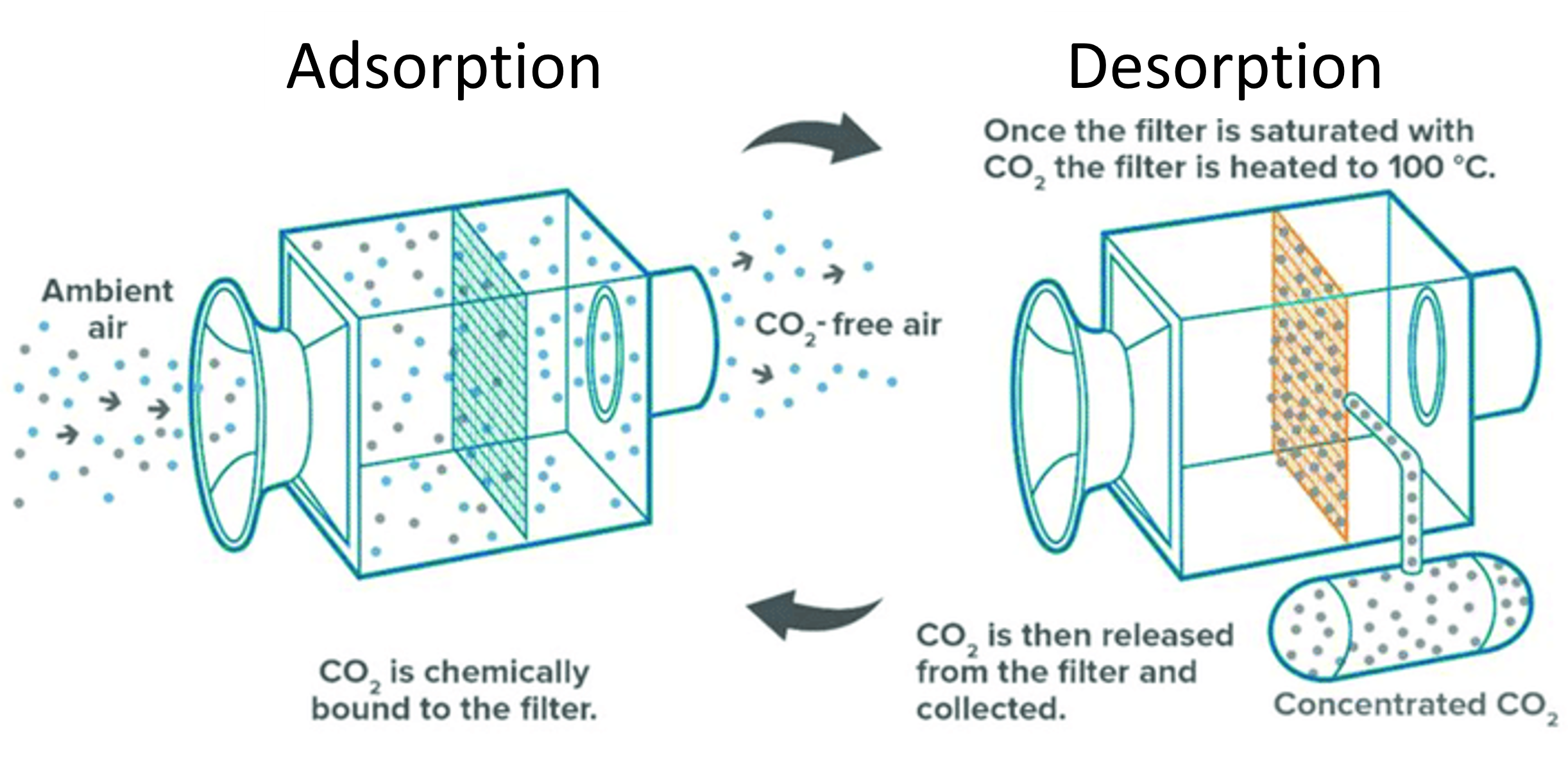}
        \hfill
        \centering 
        \includegraphics[width=0.43\textwidth]{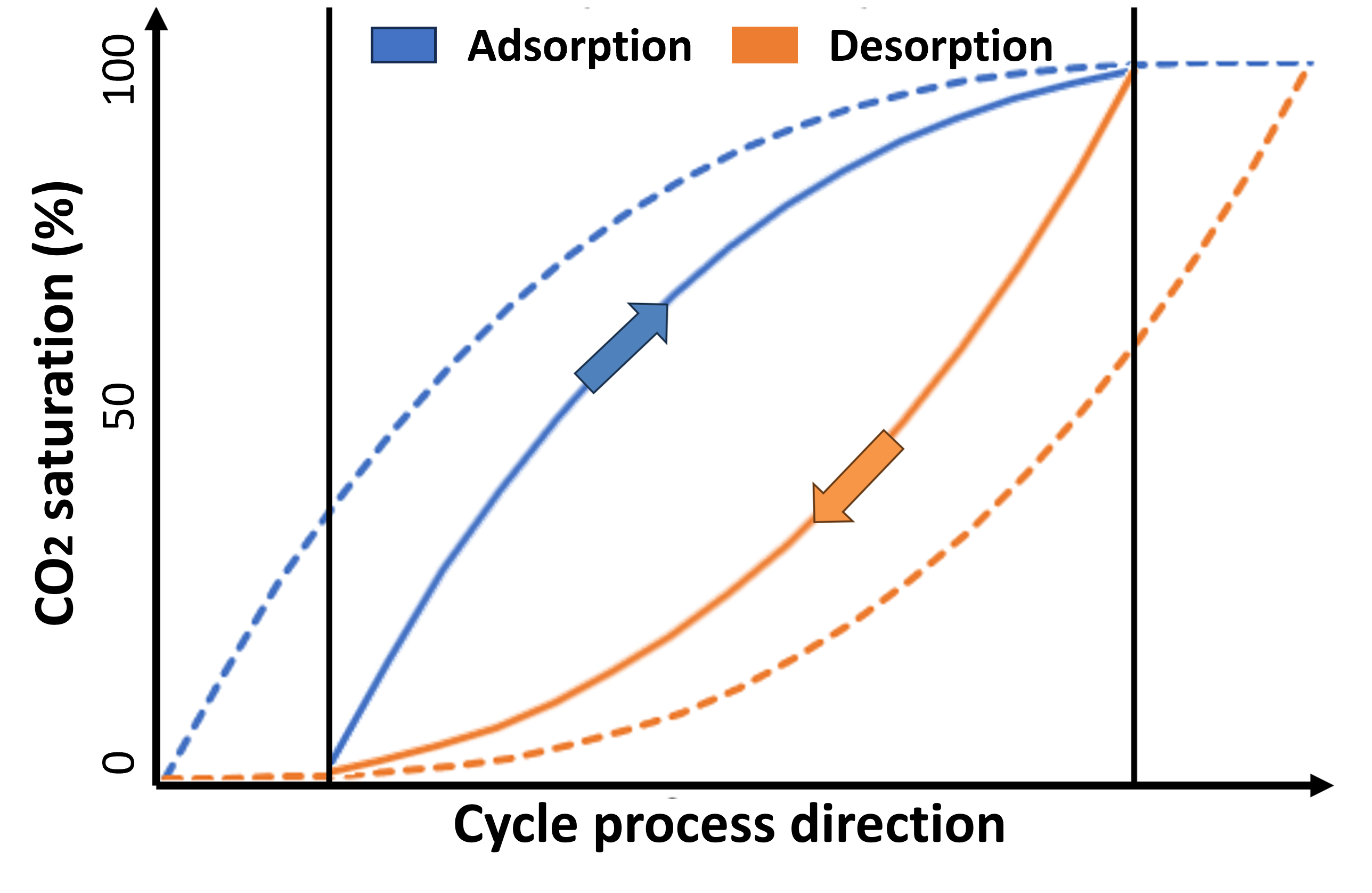} 
        \caption{\textbf{DAC system schematic and process explanation.} (left) Temperature swing adsorption DAC system schematic, where sorbent material is cycled between adsorption and desorption of \cd. During adsorption, ambient air is filtered through the sorbents, releasing \cd-lean air. In the desorption phase, captured \co is released from sorbents for storage or utilization \cite{beuttler_role_2019}. (right) The rates of adsorption and desorption are frequently non-linear with the state of \co saturation, where a high \co saturation level will make adsorption slower but desorption faster, a low \co saturation level inversely. The dashed line denotes the full-cycle DAC adsorption-desorption operation, while the solid line illustrates potential accelerated operation, which entails a minor sacrifice in \co removal per cycle but significantly reduces the overall time.}
        \label{tab:fig1}
\end{figure*}

Direct air capture (DAC) is among the most scalable technologies for significant \co removal~\cite{goldberg_co-location_2013, mcqueen_review_2021}. It extracts \co from ambient air~(\textcolor{blue}{Figure~\ref{tab:fig1}}), while captured \co can be stored geologically or mineralized for permanent removal (\(>\)10,000 years) \cite{alcalde_estimating_2018}, or utilized as a carbon feedstock for synthetic fuels \cite{parigi_power--fuels_2019}, chemicals \cite{gulzar_carbon_2020}, and building materials \cite{liu_new_2021} to support a circular carbon economy. Among different DAC design concepts, adsorption-based design using solid sorbents offers the advantages of low regeneration temperature (\(<\)100\textdegree C)\cite{mcqueen_review_2021, fasihi_techno-economic_2019, sabatino_comparative_2021}, flexibility with cyclic operation \cite{wiegner_optimal_2022} \cite{jiang_sorption_2023}, and modular design for scalability \cite{ozkan_current_2022, smith_biophysical_2016}. During the adsorption and desorption cycles for DAC system regeneration, the rates of \co capture and release exhibit nonlinearity depending on sorbent's state of saturation and the cycle time is highly sensitive to different sorbent materials and operational decisions (\textcolor{blue}{Figure~\ref{tab:fig1}}). This allows for optimization opportunities in the DAC sorbent selection and operations and potential to achieve greater net \co removal. Also, compared to liquid solvents like potassium hydroxide (KOH), solid sorbents provide wider design flexibility, allowing for potential customization of sorbent material properties, such as cycle time and temperature preferences \cite{bose_challenges_2024,young_process-informed_2023}.

\begin{table*}[!ht]
\footnotesize\rmfamily
  \centering
  \begin{threeparttable}
  \caption{\rmfamily Comparison of existing literature on DAC economics and performance modeling.}
  \setlength{\tabcolsep}{0.4mm}{
      \begin{tabular}{l | p{7cm} |
     >{\centering\arraybackslash}p{1.8cm} |
     >{\centering\arraybackslash}p{2.2cm} |
     >{\centering\arraybackslash}p{1.8cm} |
     >{\centering\arraybackslash}p{1.8cm} }
    \toprule
    Reference & Methods & Electricity Price Volatility & LCA and Emission Tracking & DAC Operator Profit & Ambient Sensitivity \\
    \midrule
    \cite{fasihi_techno-economic_2019} & Techno-economic analysis for both high-temp and low-temp DACs including fully electrified scenarios & $\times$ & $\times$ & $\Delta$ & $\times$ \\
    \hline
    \cite{sabatino_comparative_2021} & Comparative techno-economic analysis with 3 DAC techs including ambient sensitivity and cost analysis & $\times$ & $\times$ & $\Delta$ & $\checkmark$ \\
    \hline
    \cite{terlouw_life_2021} & Global Life cycle assessment (LCA) of DAC with multiple layouts with ambient temp inclusion for best DAC siting with maximum climate impact & $\times$ & $\checkmark$ & $\times$ & $\checkmark$ \\
    \hline
    \cite{breyer_carbon_2020} & Cost-optimised DAC+energy analysis using full hourly \& high spatial resolution energy system transition modeling & $\times$ & $\checkmark$ & $\Delta$ & $\times$ \\
    \hline
    \cite{daniel_techno-economic_2022} & Techno-economic analysis of DAC with \co utilisation (including net present value and levelized costs) & $\times$ & $\times$ & $\checkmark$ & $\times$ \\
    \hline
    \cite{realmonte_inter-model_2019} & Inter-model assessment using two IAMs to explore DACCS deployment scenarios and mitigation impacts & $\times$ & $\checkmark$ & $\Delta$ & $\times$ \\
    \hline
    \cite{arwa_impact_2025} & Generation capacity expansion optimization model integrating flexible DAC operations and ambient condition response & \textasciitilde & $\times$ & $\Delta$ & $\checkmark$ \\
    \hline \\
    This Paper & Profit-max DAC operation optimization with high temporal resolution data on electricity price volatility, carbon intensity and ambient sensitivity across multiple DAC techs in multiple sites/markets with respect to different policy incentives. & $\checkmark$ & $\checkmark$ & $\checkmark$ & $\checkmark$ \\
    \bottomrule
    \end{tabular}%
    }\label{literature review}
    \begin{tablenotes}
\item[a] \textasciitilde means DAC operation flexibility is added but not dynamically respond to electricity price volatility.
\item[b] $\Delta$ means techno-economic analysis of average DAC cost is tracked but the detailed financial performance of DAC system are not presented, such as (1) dynamic operation for profit maximization; (2) profit and cast flow with respect to given policy; (3) investment return period. 
    \end{tablenotes}
  \end{threeparttable}\vspace{-0.5cm}
\end{table*}

While technological advancements and scalability of DAC systems are progressing rapidly, the challenges related to its economic efficiency and energy supply have become increasingly pronounced. Every Gt of \co removal using DAC requires 167-305 TWh of electricity (even after future technical improvement), potentially increase by 6-7 times with heating electrification, and costing hundreds of billion dollars \cite{mcqueen_review_2021, fasihi_techno-economic_2019, realmonte_inter-model_2019}. In comparison, 1\% of global total electricity generation in 2021 is 280 TWh \cite{iea_world_2022} and 1\% of global GDP in 2021 is 968 billion dollars \cite{world_bank_databank_2023}. Such extensive power consumption is unlikely to be operated outside the supply-demand balance in power market, whose costs rely solely on subsidies funded by taxpayers. Due to its enormous potential scale, DAC’s energy demand could place a significant burden on the economy — particularly if it is planned or operated inefficiently, leading to both economic losses and additional GHG emissions from energy use. A recent study \cite{herzog_getting_2024} calls for a more realistic analysis for commercial CDR deployment, considering scaling, energy, location, and cost in more practical ways.

\subsection{Literature review and research gap}

Previous studies on the economic efficiency and GHG life cycle assessment (LCA) of DAC have shown that a grid-connected system is sensitive to the fossil electricity mixture, greatly affecting its capture efficiency and potentially lead to net \co emissions rather than removal \cite{terlouw_life_2021, deutz_life-cycle_2021}. These studies are generally based on \textit{average electricity} price and emission intensity assumptions, seldom include renewable profiles \cite{breyer_carbon_2020}, and not considering temporal volatility in electricity price and emission as shown in \textcolor{blue}{Table 1}. These assumptions are based on constant DAC operations - approximately 100\% capacity factor- and thus exclude potential operational optimization. Typically, techno-economic analyses focus on minimizing system costs to achieve specific CDR targets without accounting for the utility of individual DAC systems as autonomous agents. However, for commercial DAC projects, it is more rational to also consider the DAC system operator standpoints, who are \textbf{profit-driven}, especially when upfront investment is substantially made through private funding sources \cite{carter_private_2020}. Commercial operations are also subjected to high price volatility introduced by large expansions in renewable power supplies \cite{maniatis_impact_2022}. A well-known example is the ``duck curve" observed in California (CA) due to high solar generation during the daytime \cite{calero_duck-curve_2022, caiso_california_2023}. The real-time price in power markets updates sub-hourly, CA and New York (NY) update the prices every 5 minutes \cite{caiso_california_2023, nyiso_new_2023}, and Texas (TX) every 15 minutes \cite{ercot_electric_2023}, and both price and emission intensity change rapidly. Intuitively, while exposed to market volatility without a minimum \co removal constraint, DAC operations may be expected to accelerate when the power price is low and reduce or shut down when the price is high. Additionally, DAC operations are sensitive to ambient temperature and relative humidity while different types of DAC systems have different preferred environments \cite{an_impact_2022}\cite{sendi_geospatial_2022}. These characteristics point to the need for site-specific DAC operation modeling with higher resolution.

\subsection{Contribution}

Based on the review over existing literature and research gap, our contribution are presented:
\begin{itemize}
    \item \textbf{Profit-driven optimization from DAC system perspective.} Our model optimizes DAC system operations with a profit-maximization objective, thereby capturing the realistic behavior of these systems under conditions of price volatility and evolving policy environments/penalties, e.g., incentives and carbon taxes.
    \item \textbf{High and flexible temporal resolution.} Our model employs a 5-minute temporal resolution, matching the electricity market price data, which can be flexibly aggregated to lower resolutions with corresponding adjustments to the DAC parameters. Our analysis reveals that certain temporal dynamics are critical and can only be observed at high temporal resolutions.
    \item \textbf{Spatial and ambient environment sensitivity.} We analyzed three representative U.S. electricity markets—New York, California, and Texas—each characterized by distinct price profiles driven by different dominant renewable energy sources and unique ambient conditions (e.g., temperature and relative humidity) that significantly influence DAC operations. Our findings indicate that optimal DAC siting may be even more critical than engineering improvements.
    \item \textbf{Reverse engineering future DAC R\&D.} Our analysis indicates that under conditions of highly volatile electricity prices, carbon intensities, and ambient environments, "flexible DAC" operation/capability is not only strictly preferred but also offers significant utility—highlighting a previously underestimated research direction for future DAC system research and development. 
\end{itemize}


\section{Models}

There are primarily two types of operations for different DAC systems: cyclic and continuous. In cyclic systems, typically low-regeneration temperature sorbents (100\textdegree C), a stationary chamber holds the sorbent material, which undergoes adsorption-desorption cycles through variations in temperature, pressure, humidity, or a combination thereof (e.g., temperature-pressure swing design). In contrast, continuous systems typically employ a material loop for high regeneration temperature solvents (800\textdegree C for CaCO\textsubscript{3} regeneration), where sorbent or solvent materials are conveyed between different pieces of equipment for each process step. Although the materials in continuous systems also experience absorption-regeneration cycles, key equipment and its energy consumption—such as adsorption platforms and regeneration units—operates continuously by constantly moving materials. This study models both processes, with a particular focus on cyclic operation due to its low-regeneration temperature and non-linear optimization characteristics. Both models are presented in rigorous mixed integer linear programming (MILP) optimization problems we propose additional \textbf{strategic bidding} algorithm that improves the computation. 

\subsection{Cyclic model}
The objective function maximizes the total profit of the DAC operation at each time step within the horizon \(t\in T\).

\begin{equation}
    max \sum_t \; \pi d_t - \lambda_t C_{t} (P^a u_t +  P^d v_t)-Sz_t
\end{equation}
where  \(\pi d_t\) is the total revenue by captured (desorbed) \co, minus the cost of power consumption corrected by ambient conditions at each time step \( \lambda_t C_{t} (P^a u_t +  P^d v_t)\), minus the cost of material consumption (sorbent degradation and replacement and fixed O\&M) for switching cycles \(Sz_t\). \rev{The fixed material replacement cost is effectively scaled proportionally with the operating capacity factor, reflecting a linear relationship of utilization on sorbent/solvent lifetime.}

Binary constraints for absorption/desorption status:

\begin{equation}
    u_t + v_t \leq 1
\end{equation}
where the DAC system can only be either adsorption or desorption at one time period, but it can be neither absorbing nor desorbing, i.e., staying idle. Absorption and desorption rate (using inequality here to avoid contradiction with binary constraints, which is relaxation of the constraints) are subjected to state-of-saturation \(\bar{X}\) correct to model the nonlinear behavior. Specifically, only adsorption (capture) process is corrected by ambient condition \(\eta^{a}_{t}\) of each time step, where desorption is a controlled high-temperature process that is not affected by ambient condition:
\begin{equation}
    a_t \leq \eta^{a}_{t} (\beta^{a}_{1} + \beta^{a}_{2} X_t)
\end{equation}
\begin{equation}
    d_t \leq \beta^{d}_{1} + \beta^{d}_{2} X_t
\end{equation}
additionally, the absorption rate and desorption rate shall be bounded by the binary variable for each as well.
\begin{equation}
    a_t \leq M u_t
\end{equation}
\begin{equation}
    d_t \leq M v_t
\end{equation}
here \(M\) is a sufficiently large number which does not bind the absorption and desorption rate if \(u_t\) and \(v_t\) are 1.
\\


\makenomenclature
\nomenclature[A]{$u_t$}{Binary variable; 1 if cyclic DAC is in the absorption phase at time period $t$, 0 otherwise.}
\nomenclature[A]{$v_t$}{Binary variable; 1 if cyclic DAC is in the desorption phase at time period $t$, 0 otherwise.}
\nomenclature[A]{$z_t$}{Binary variable; 1 if cyclic DAC switches to a new cycle at time period $t$, 0 otherwise.}
\nomenclature[A]{$k_t$}{Binary sign-variable to facilitate calculation of $z_t$ by determining change of status at time period $t$.}
\nomenclature[A]{$w_t$}{Binary variable for continous DAC; 1 if DAC is operating at time period $t$, 0 otherwise}
\nomenclature[B]{$X_t$}{Continuous non-negative state-of-saturation capacity of the DAC system at time period $t$.}
\nomenclature[B]{$a_t$}{Continuous non-negative absorption amount of the DAC system at time period $t$.}
\nomenclature[B]{$d_t$}{Continuous non-negative desorption amount of the DAC system at time period $t$.}

\nomenclature[C]{$\lambda_t$}{Electricity price at time period $t$, potentially modified by $CO_2$ intensity.}
\nomenclature[C]{$e_t$}{Electricity $CO_2$-intensity at time period $t$.}
\nomenclature[C]{$\rho_e$}{$CO_2$ value of carbon for electricity; equals 0 for wholesale electricity price or $\pi$ for carbon-tax adjusted electricity price.}
\nomenclature[C]{$\lambda e_t$}{Electricity price without $CO_2$-intensity correction (i.e., $\lambda_t = \lambda e_t + \rho_e e_t$).}
\nomenclature[C]{$\eta_t$}{Adsorption rate correction factor at time period $t$, as a function of relative humidity and temperature.}
\nomenclature[C]{$C_t$}{Energy consumption correction factor at time period $t$, as a function of relative humidity and temperature.}
\nomenclature[C]{$\pi$}{$CO_2$ value constant (includes selling price, subsidies, carbon tax, etc.).}
\nomenclature[C]{$M$}{A general large constant that is used to bound the continuous decision variables with associated binary status.}

\nomenclature[D]{$P^a$}{Electricity consumption for the absorption phase per unit time period.}
\nomenclature[D]{$P^d$}{Electricity consumption for the desorption phase per unit time period.}
\nomenclature[D]{$\bar{X}$}{Maximum available DAC capacity.}
\nomenclature[D]{$S$}{Switching cycle cost for the consumption of sorbent materials.}

\nomenclature[D]{$\beta^a_1$}{First-order coefficient for absorption.}
\nomenclature[D]{$\beta^a_2$}{Second-order coefficient for absorption.}
\nomenclature[D]{$\beta^d_1$}{First-order coefficient for desorption.}
\nomenclature[D]{$\beta^d_2$}{Second-order coefficient for desorption.}

\nomenclature[N]{DAC}{Direct air capture.}
\nomenclature[N]{CDR}{Carbon dioxide removal.}
\nomenclature[N]{O\&M}{Operation and Maintenance.}
\twocolumn[{%
    \begin{center}
    \end{center}
    \begin{framed}
    \begin{multicols}{2}
    \printnomenclature
    \end{multicols}
    \end{framed}
}]


State-of-saturation capacity of DAC system updates with absorption/desorption rates:
\begin{equation}
    X_t-X_{t-1} = a_t - d_t
\end{equation}
where the change of state-of-saturation between time periods equals: adding absorption or subtracting desorption.
\\
\begin{equation}
    0 \leq X_t \leq \bar{X}
\end{equation}
where the state-of-saturation is always lower bounded by 0 (non-negative), and upper bounded by declared maximum capacity \(\bar{X}\).
\\
Switching cycle constraints model the cost of each cycle, this includes cost of degradation of sorbent materials and fixed cost for regeneration heat supply. The initial state of the sign-variable equals indicating cycle switching is zero.
\begin{equation}
    k_0 = 0
\end{equation}
The following constraints allocate idle states as continueing of the previous adsorption/desorption process so idle states won't triger cycle counting:

\begin{equation}
    -M(1-k_t) \leq k_{t-1} + (u_t - v_t) - 0.5
\end{equation}
\begin{equation}
    M k_t \geq k_{t-1} + (u_t - v_t) - 0.5
\end{equation}
The solution matrix for \(k_t\) cross all possible combinations of sign-function is presented in the table below. It logically interprets the idle status as continuation of the ongoing adsorption/desorption process, rather than triggering a new cycle that consume material lifespan and iccur cost:
\begin{center}
\begin{tabular}{||c c c||} 
 \hline
 Variables & \(k_{t-1} = 0\) & \(k_{t-1} = 1\)\\ [0.5ex] 
 \hline\hline
 \(u_t-v_t\) = -1 & 0 & 0 \\
 \hline
 \(u_t-v_t\) = 0 & 0 & 1 \\
 \hline
 \(u_t-v_t\) = 1 & 1 & 1 \\ [1ex] 
 \hline
\end{tabular}
\end{center}

Then the sign binary variable \(k_t\) can be used to determine the binary cycle counting variable \(z_t\):
\begin{equation}
    z_t \geq k_t - k_{t-1}
\end{equation}
where \(z_t\) will be minimized in the objective function that only when \(k_t = 1\) and \(k_{t-1}=0\), \(z_t = 1\).

The above formulation summarizes the optimization framework of the DAC system using the temperature-pressure-swing DAC technology (all 3 cyclic DAC technologies tested in this paper). In practice, the DAC input parameters (\(\beta\) values) including the piecewise linear approximation for absorption and desorption rate using quadratic coefficients will be determined by different DAC specifications. \rev{For cyclic DAC systems, which naturally alternate between adsorption and desorption modes, minimum stable load, ramping rates, and on/off cycle limits are not applicable.}

\subsection{Continuous model}
The variables definition is consistent with the cyclic DAC MILP settings.

\begin{equation}
    max \sum_t \; \pi d_t \eta^{a}_{t} - \lambda_t P_t^d - S_c d_t
\end{equation}
where  \(\pi d_t\) is the total revenue corrected by ambient environment conditions \(\eta^{a}_{t}\) minus the cost of power consumption \(\lambda_t(P^a u_t + P^d v_t)\) and operational cost \(S_c d_t\) which include both thermal energy cost and material consumption. The correction of ambient condition is applied on desorption directly as a continuous process is assumed to seemlessly desorb the \co instantly after adsorption.

For continuous DAC, the operational flexiblity is determined by two constraints. The first one is ON/OFF limit constraints:

\begin{equation}
    y_t - z_t = w_t - w_{t-1}
\end{equation}
\begin{equation}
    y_t - z_t \leq 1
\end{equation}
where \(w_t\) is the binary on/off status of DAC, \(y_t\) and \(z_t\) indicating switching on and off status of DAC.
\begin{equation}
    \sum_{\tau = 1}^I (y_{\tau} + z_{\tau}) \leq K
\end{equation}
meaning the total times of switching on or off during the horizon (\(i \in 1,2,...,I\)) is limited by maximum flexibility index \(K\). In practice, \(I = 288=24*(60/5)\) 5-min time-steps leading to 1 day horizon, meaning the continuous DAC can maximum turn on/off by \(K\) times in one day.

The second one is min/max rate constraints:
\begin{equation}
    \underline{d} w_t \leq d_t \leq \bar{d} w_t
\end{equation}
the capture rate of \co \(d_t\) is limited between min/max capture rate, and we use 80\% to 100\% as presentative knowing this might vary for different technologies. Accompanied with the power consumption constraints:
\begin{equation}
    P_t^d = \frac{d_t}{\bar{d}} \bar{P^t}
\end{equation}
where the actual power consumption \(P_t^d\) is the fraction of nominal power consumption for capture \(\bar{P^t}\).

\subsection{Abatement Accounting}
\rev{Eventually, the net-CO2 removal is calculated by summing the total desorption and subtracting the emission from power consumption: emission intensity et times power consumption.}
\begin{equation}
    CO_2^{net} = \sum_t{d_t} - e_t(P^a u_t + P^d v_t)
\end{equation}
\rev{\co capture efficiency is defined by net-\co removal over total captured:}
\begin{equation}
    \eta_{CO_2} = \frac{CO_2^{net}}{\sum_t{d_t}}
\end{equation}

\subsection{Strategic bidding algorithm}

\begin{figure*}[!ht]
        \centering
        \includegraphics[width=0.80\textwidth]{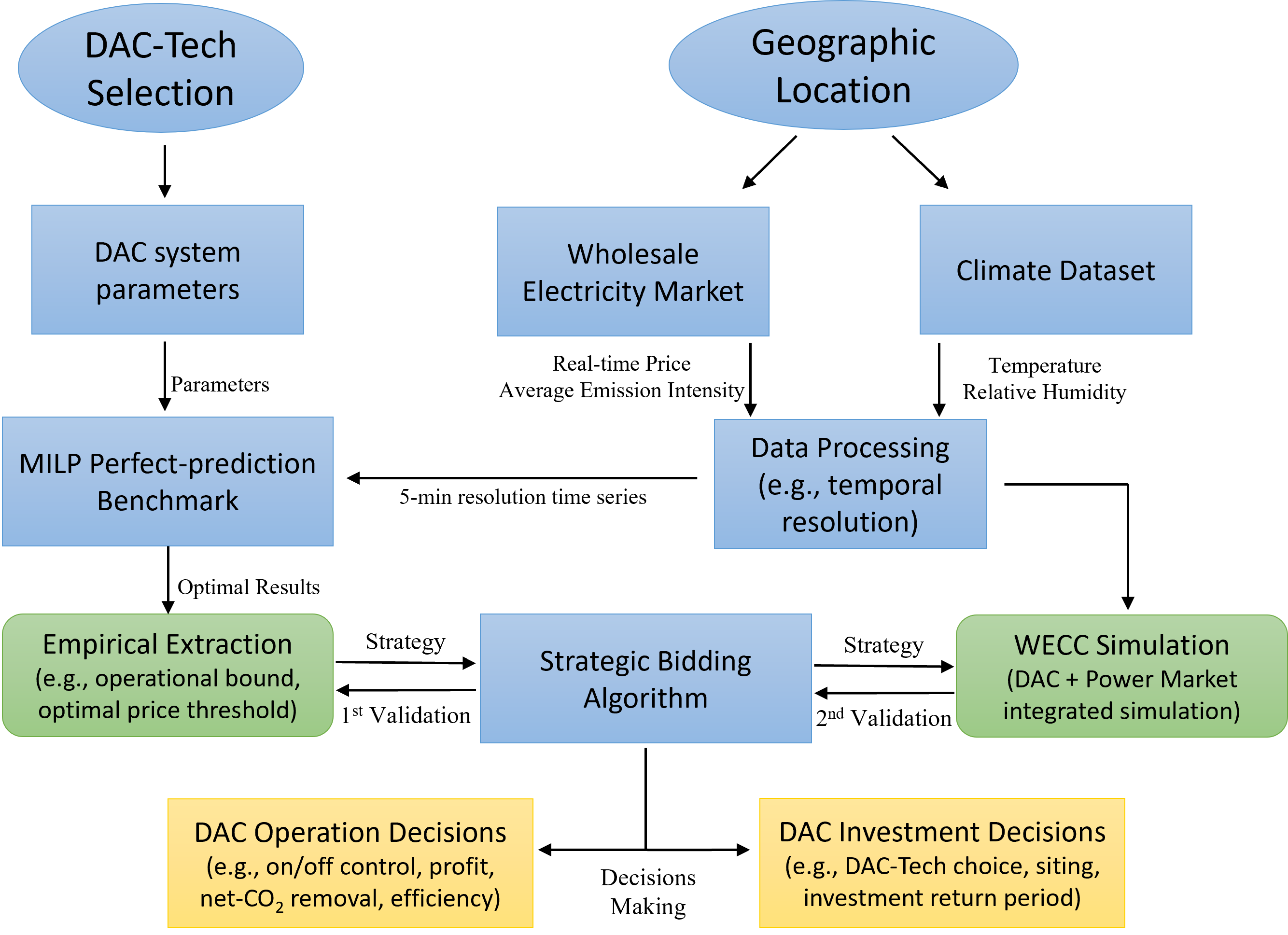}
        \caption{\textbf{Modeling Framework: Data and process flow schematic.} The schematic shows data/parameter sources and selection, preprocessing steps, validation and results for the DAC–power market analysis.}
        \label{tab:schematic}
\end{figure*}

The computational burden of both DAC MILP models are expensive, where a 1-year rolling-horizon optimization generally takes 4-8 hours for cyclic model and 1.5-2 hours for continuous model. We propose a "strategic bidding algorithm" for cyclic DAC operation that improves the computational efficiency and simulate the actual DAC system behavior. \rev{Based on the MILP benchmark using deterministic price scenarios, we empirically extract operational insights such as bidding thresholds based on market clearing horizon (typically 1 day) and operate with the following logic:}
\[
\begin{cases} 
\text{Activate} & \text{if } \lambda_t(t)\leq \lambda_{\text{opt}} \\
\text{Idle} & \text{otherwise}
\end{cases}
\]
when active during low-price period, the DAC system will periodically adopt adsorption-desorption cycles between the pre-determined depth:
\begin{equation}
    c_1\bar{X} \leq X_t \leq c_2\bar{X}
\end{equation}
where \(c_1\) and \(c_2\) can be learned statistically from observing rigorous MILP results. \rev{See \textcolor{blue}{Figure}~\ref{tab:schematic} for the workflow.}

The proposed strategic bidding algorithm offers several advantages (quantitative details see \textcolor{blue}{Table~\ref{tab:milp_bidding}}):

\begin{itemize}
    \item \textbf{Realistic bidding strategy:} It simulates a scenario in which a DAC system submits a demand bid to a power system operator, expressed as a price threshold: if the actual electricity price exceeds the bid, the system remains idle; if the price falls below the threshold, the DAC system operates. This is a more realistic market representation than the theoretical MILP optimum.
    \item \textbf{Near-optimal performance:} It exhibits near-optimal performance—approximately 90\% of rigorous MILP results in most cases—and maintains stable solutions even under negative or marginal profit scenarios
    \item \textbf{Interpretability:} It provides clear and simple operational guidelines for the DAC system by defining electricity price cut-offs and identifying the optimal state-of-saturation depth for the specified technologies.
    \item \textbf{Computational efficiency:} It enhances computational speed by a factor of 1,500–3,000, reducing the optimization process to approximately 10 seconds for a one-year optimization. Essentially, the optimization problem is reduced to a single variable, \(\lambda_{\text{opt}}\), as opposed to managing hundreds of variables, including binary ones.
\end{itemize}

\begin{table}[H]
\centering
\caption{Representative one-week comparison of the MILP benchmark and the custom bidding strategy for MOF operation in California.}
\label{tab:milp_bidding}
\resizebox{\columnwidth}{!}{%
\begin{tabular}{lccccc}
\hline
\textbf{Method} & \textbf{Run-} & \textbf{Profit} & \textbf{Removal} & \textbf{Removal} & \textbf{Cycles} \\
 & \textbf{time} (s) & (\$) & \textbf{Tot.} (t) & \textbf{Net} (t) & \\
\hline
MILP    & 566   & 1589 & 80.1 & 59.2 & 101 \\
Bidding & 0.23  & 1550 & 86.3 & 66.3 & 108 \\
\hline
\end{tabular}%
}
\end{table}

\rev{To address potential feedback effects of DAC operations on market prices, we provide a comparison between the “price taker” assumption presented here and “price influencer” cases using integrated DAC + WECC power market simulation results. We find that from the kiloton to hundred-kiloton scale, these feedback effects remain manageable, while strong feedback appears when DAC deployment reaches the scale with GW level peak load (Mt-\cd/year scale or higher).}



\section{Trading-off \co removal with economic efficiency}\label{sec2}

\begin{table*}[h!]
    \centering
    \caption{Comparison of profit-maximization operation and full-capacity operation with different DAC technologies}
    \begin{tabular}{r|ccccccccccccc}
        \toprule
        \textbf{DACs} & \multicolumn{3}{c|}{\textbf{MOF}} & \multicolumn{3}{c|}{\textbf{APDES-NFC-FD}} & \multicolumn{3}{c|}{\textbf{SI-AEATPMS}} & \multicolumn{3}{c}{\textbf{KOH}} \\
        \midrule
        \textbf{Cycle time (hours)} & \multicolumn{3}{c|}{1.46} & \multicolumn{3}{c|}{9.53} & \multicolumn{3}{c|}{89.2} & \multicolumn{3}{c}{continuous} \\
        \midrule
        \multicolumn{13}{c}{\textbf{Energy Consumption} (MWh per ton-\co captured)}\\
        \midrule
        \textbf{Electricity - Adsorption} & \multicolumn{3}{c|}{0.642} & \multicolumn{3}{c|}{0.300} & \multicolumn{3}{c|}{0.357} & \multicolumn{3}{c}{0.300} \\
        \textbf{Electricity - Desorption} & \multicolumn{3}{c|}{0.097} & \multicolumn{3}{c|}{0.060} & \multicolumn{3}{c|}{0.071} & \multicolumn{3}{c}{0.200} \\
        \textbf{Thermal} & \multicolumn{3}{c|}{3.15} & \multicolumn{3}{c|}{1.43} & \multicolumn{3}{c|}{1.89} & \multicolumn{3}{c}{2.25} \\
        \midrule
        \multicolumn{13}{c}{\textbf{Cycle cost/OPEX excluding electricity cost} (\$ per ton-\co captured)}\\
        \midrule
        \textbf{Thermal} & \multicolumn{3}{c|}{75.60} & \multicolumn{3}{c|}{34.32} & \multicolumn{3}{c|}{46.36} & \multicolumn{3}{c}{54.00} \\
        \textbf{Sorbent/solvent Material} & \multicolumn{3}{c|}{40.00} & \multicolumn{3}{c|}{7.70} & \multicolumn{3}{c|}{168.00} & \multicolumn{3}{c}{20.00} \\
        \textbf{Total} & \multicolumn{3}{c|}{115.6} & \multicolumn{3}{c|}{42.02} & \multicolumn{3}{c|}{213.36} & \multicolumn{3}{c}{74.00} \\
        \midrule
        \multicolumn{13}{c}{\textbf{Same investment plants}}\\
        \midrule
        \textbf{CAPEX (\$/ton-\cd-yr)} & \multicolumn{3}{c|}{88} & \multicolumn{3}{c|}{214} & \multicolumn{3}{c|}{221} & \multicolumn{3}{c}{609} \\
        \textbf{Investment (million \$)} & \multicolumn{3}{c|}{10.50} & \multicolumn{3}{c|}{10.50} & \multicolumn{3}{c|}{10.50} & \multicolumn{3}{c}{10.50} \\
        \textbf{Lifetime (years)} & \multicolumn{3}{c|}{20} & \multicolumn{3}{c|}{20} & \multicolumn{3}{c|}{20} & \multicolumn{3}{c}{20} \\
        \textbf{Capacity (ton-\cd/yr)} & \multicolumn{3}{c|}{6000} & \multicolumn{3}{c|}{2453} & \multicolumn{3}{c|}{2375} & \multicolumn{3}{c}{862} \\
        \midrule
        \textbf{Results} & \multicolumn{3}{c|}{\textbf{MOF}} & \multicolumn{3}{c|}{\textbf{APDES-NFC-FD}} & \multicolumn{3}{c|}{\textbf{SI-AEATPMS}} & \multicolumn{3}{c}{\textbf{KOH}} \\
        \midrule
        \textbf{Full capacity} & \multicolumn{1}{c|}{NY} & \multicolumn{1}{c|}{CA} & \multicolumn{1}{c|}{TX} &\multicolumn{1}{c|}{NY} & \multicolumn{1}{c|}{CA} &\multicolumn{1}{c|}{TX} & \multicolumn{1}{c|}{NY} & \multicolumn{1}{c|}{CA}& \multicolumn{1}{c|}{TX} & \multicolumn{1}{c|}{NY} & \multicolumn{1}{c|}{CA} & \multicolumn{1}{c}{TX}\\
        \midrule
        \textbf{Capacity factor} & 100\% & 100\% & 100\% & 100\% & 100\% & 100\% & 100\% & 100\% & 100\% & 100\% & 100\% & 100\% \\
        \textbf{Profit (1000\(\$\))} & -5.2 & -65.0& 13.1 & 333.9 & 325.8 &337.3 & -100.9 & -108.6 &-97.3 & 82.8 & 79.2 & 84.8\\
        \textbf{Net-\co removal (ton/yr)} & 4,434 & 4,553 &4,104 & 2,205 & 2,224 & 2,152& 2,076 & 2,099 &2,016 & 738 & 741 & 712 \\
        \textbf{\co capture efficiency} & 74\% & 76\% & 68\%& 90\% & 91\% & 88\% & 88\% & 89\% &85\% & 86\% & 87\% & 83\% \\
        \midrule
        \textbf{Profit-driven} & \multicolumn{1}{c|}{NY} & \multicolumn{1}{c|}{CA} & \multicolumn{1}{c|}{TX} &\multicolumn{1}{c|}{NY} & \multicolumn{1}{c|}{CA} &\multicolumn{1}{c|}{TX} & \multicolumn{1}{c|}{NY} & \multicolumn{1}{c|}{CA}& \multicolumn{1}{c|}{TX} & \multicolumn{1}{c|}{NY} & \multicolumn{1}{c|}{CA} & \multicolumn{1}{c}{TX}\\
        \midrule
        \textbf{Capacity factor} & 64\% & 50\% & 70\%& 98\% & 98\% &98\% & 0\% & 9\% &0\% & 99\% & 94\% & 98\% \\
        \textbf{Profit (1000\(\$\))} & 94.9 & 153.5 &127.3 & 333.7 & 327.5 & 342.2& 0 & 2.5 &0 & 83.9 & 82.2 & 88.3\\
        \textbf{Net-\co removal (ton/yr)} & 2,919 & 2,408 & 2,982& 2,165 & 2,177 &2,112 & 0 & 197 &0 & 731 & 707 & 703\\
        \textbf{\co capture efficiency} & 76\% & 80\% &70\% & 90\% & 91\% & 88\%& N/A & 92\% &N/A & 86\% & 87\% & 83\% \\
    \end{tabular}
    \begin{tablenotes}
        \small
        \item *Note: Comparisons among various DAC technologies are grounded in a uniform initial investment of \$10.5 million per plant, with plant capacity (ton-\cd/yr) derived from CAPEX and lifespan considerations, excluding financial assumptions. Thermal energy consumption is fixed, whose cost is included in cycle cost without modeling the temporal volatility. All results are 2022 full-year cumulative with respect to the power market with incentive selling price = \$200/ton-\cd. Power markets properties: Average price: CA \$70.0/MWh; NY \$60.8/MWh; TX \$57.5/MWh. Average electricity \cd-intensity: CA 0.26 ton-\cd/MWh; NY 0.28 ton-\cd/MWh; TX 0.35 ton-\cd/MWh. Dominant low-carbon power: CA solar; NY hydropower and nuclear; TX wind. All \co removal results in this study are net-removal, meaning electricity emissions are subtracted and capture efficiency is reported.
    \end{tablenotes}
    \label{tab:ben}
\end{table*}

Our objective is to compare commercially deployable DAC technologies utilizing electricity as the primary energy source. We have chosen three sorbent DAC technologies and KOH liquid solvent system for comparison for their technical maturity and data availability (especially in cycle time and operational flexibility): 
\begin{itemize}
    \item \textbf{MOF}: Metal-Organic Framework, a family of novel solid porous adsorbents with fast cyclic operation, and potentially low CAPEX. Energy consumption is relatively high but rapidly improving through innovation. Challenges include sorbent material costs and unknown lifespan for synthesis on a large scale \cite{azarabadi_sorbent-focused_2019, sinha_systems_2017}.
    \item \textbf{APDES-NFC-FD (AN)}: a type of commercialized amine-functionalized sorbent, cyclic operation. Very efficient in energy and material cost with much less sorbent consumption per ton \co captured. Higher CAPEX, moderate flexibility in cycle time \cite{leonzio_environmental_2022, wurzbacher_heat_2016}.
    \item \textbf{SI-AEATPMS (SA)}: a type of early commercialized amine-functionalized sorbent, cyclic operation with very high sorbent consumption per ton \co capture. Expensive and inflexible operations with long cycle time. Used as a comparison benchmark \cite{leonzio_environmental_2022, marinic_direct_2023}.
    \item \textbf{KOH liquid solvent}: a commercialized continuous operation DAC by looping KOH liquid solvent. The regeneration temperature is high at about 800 \(^{\circ}\)C. Expensive in CAPEX and energy intensive for high temperature solvent regeneration \cite{sabatino_comparative_2021}\cite{keith_process_2018}.
\end{itemize}

All cyclic solid sorbent DAC technologies require nearly the same desorption-regeneration temperature (100 \(^{\circ}\)C), but larger differences in costs, cycle times, and energy consumption: 
\textcolor{blue}{Table}~\ref{tab:ben} shows the technical parameters of the four DAC technologies and full-year baseline simulation results for three selected power markets. In this power system-centric study, we assume that, aside from cost considerations, no operational constraints or geographic differentiation on thermal energy consumptions are applied. We simulate each technology using 2022 electricity prices from the NY, CA, and TX power markets (NYISO, CAISO, and ERCOT respectively), represent the three grid interconnection areas in the U.S.: Eastern, Western, and Texas. \rev{We use average emission signal published by ISOs for optimization as these are used as the primary indicator for measuring carbon emission intensity~\cite{schafer_towards_2024}.}

\begin{figure*}[h]
    \centering
        \includegraphics[width=1\textwidth]{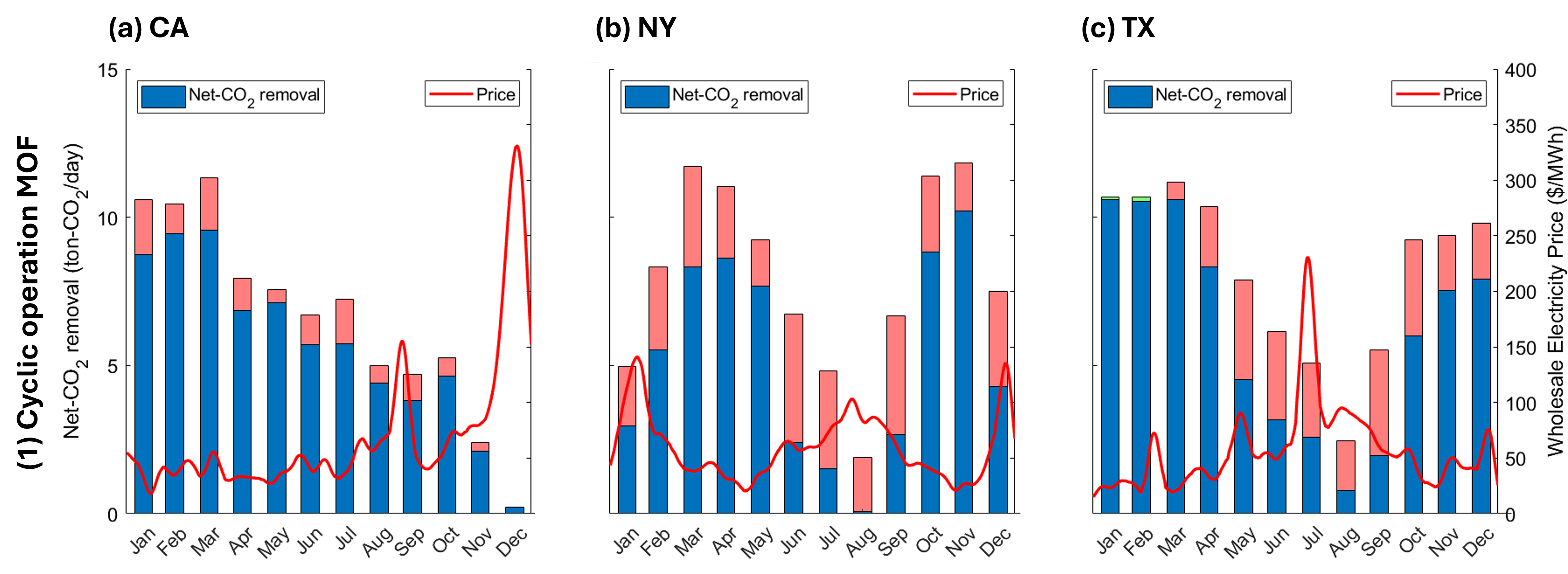}
        \includegraphics[width=1\textwidth]{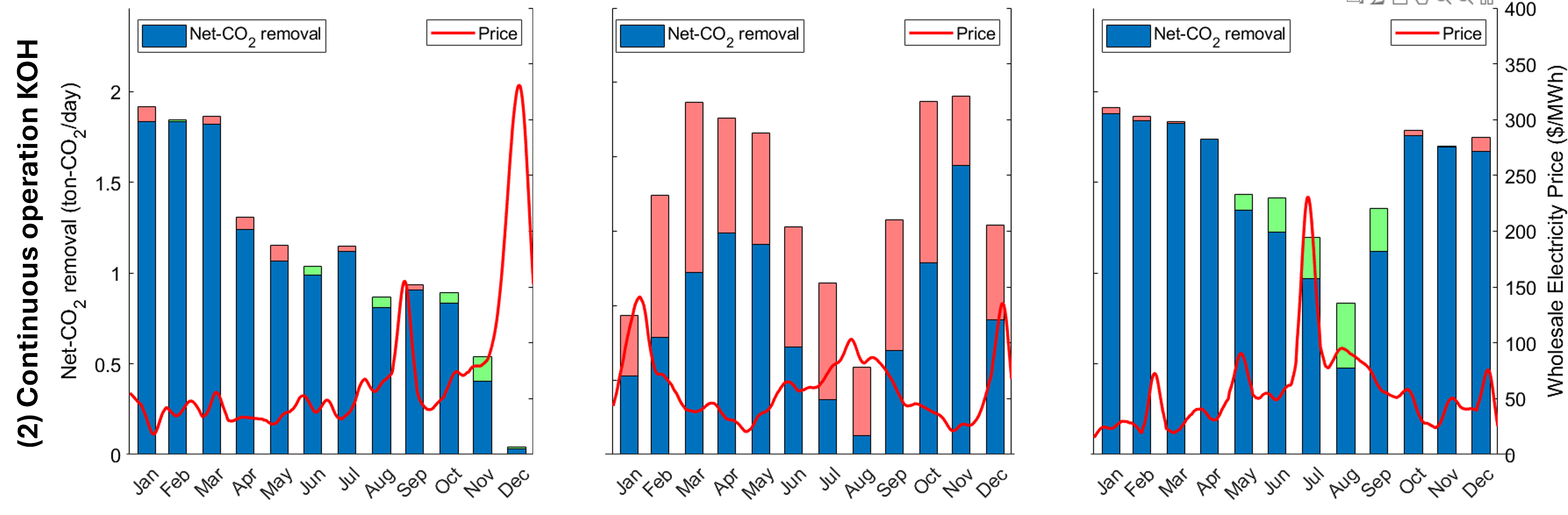} 
    \caption{\textbf{Annual wholesale electricity price profile and monthly DAC net-\co removal.} Columns: (a) CA; (b) NY; (c) TX. Rows: (1) MOF; (2) KOH. \textcolor{red}{Red} and \textcolor{green}{green} bar represents \co abatement \textcolor{red}{loss} and \textcolor{green}{gain} due to ambient temperature and relative humidity change compared to baseline lab environment shown as the \textcolor{blue}{blue} bar. The actual abatement volume is \textcolor{blue}{blue} +\textcolor{green}{green} or -\textcolor{red}{red} bars with gap as large as 30\%. \rev{Incentive selling prices of \$200/ton–\co for MOF and \$100/ton–\co for KOH are chosen such that they equal the sum of non-electricity operating costs and the product of each system’s bidding threshold with its power consumption, making both technologies marginally profitable.} \rev{This comparison shows that, despite their different cost structures and modeling approaches, different DAC technologies exhibit qualitatively similar dispatch behavior in response to electricity price profiles when operating near their marginal profitability threshold.}  Net-\co removal is assessed using daily averages within each month.}
    \label{tab:fig2}
\end{figure*}

The comparison between profit-driven and full-capacity models for different DAC technologies unveils a non-intuitive outcome: increasing \co removal does not always translate into higher profits for DAC operators. MOF technology offers more \co removal for larger plant capacities but is less profitable than AN technology. A profit-driven MOF technology offers the flexibility to avoid high-price periods, reversing non-profitable NY and CA cases profitable, 
 and enhances economic efficiency in TX by 10 times. Profitability enables earlier deployment within each power market and policy framework. DAC operations may benefit by remaining idle and reducing their \co removal efforts during certain periods. At full capacity, DAC systems could be unprofitable in the same market and under the same incentives, making such projects unviable. Investors may opt for more efficient and profitable technologies with smaller plant capacity and operate it profit-driven which reduces \co removal. The outcome points to an economic-climate trade-off: from DAC system perspective, increasing economic efficiency may reduce net \co removal.

\section{Power Market and Environmental Implications}\label{sec3}

\subsection{Annual Temporal Behaviors}\label{subsec2}

\begin{figure*}[h!]
    \centering
        \includegraphics[width=1\textwidth]{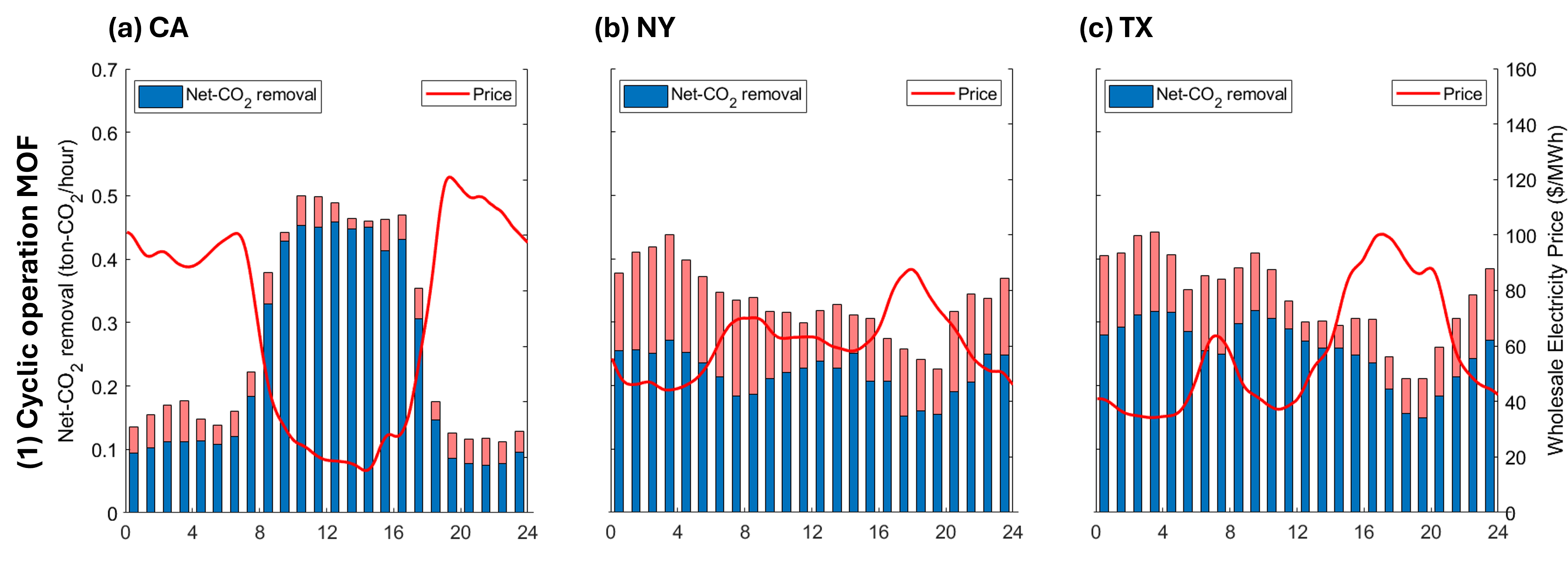}
        \includegraphics[width=1\textwidth]{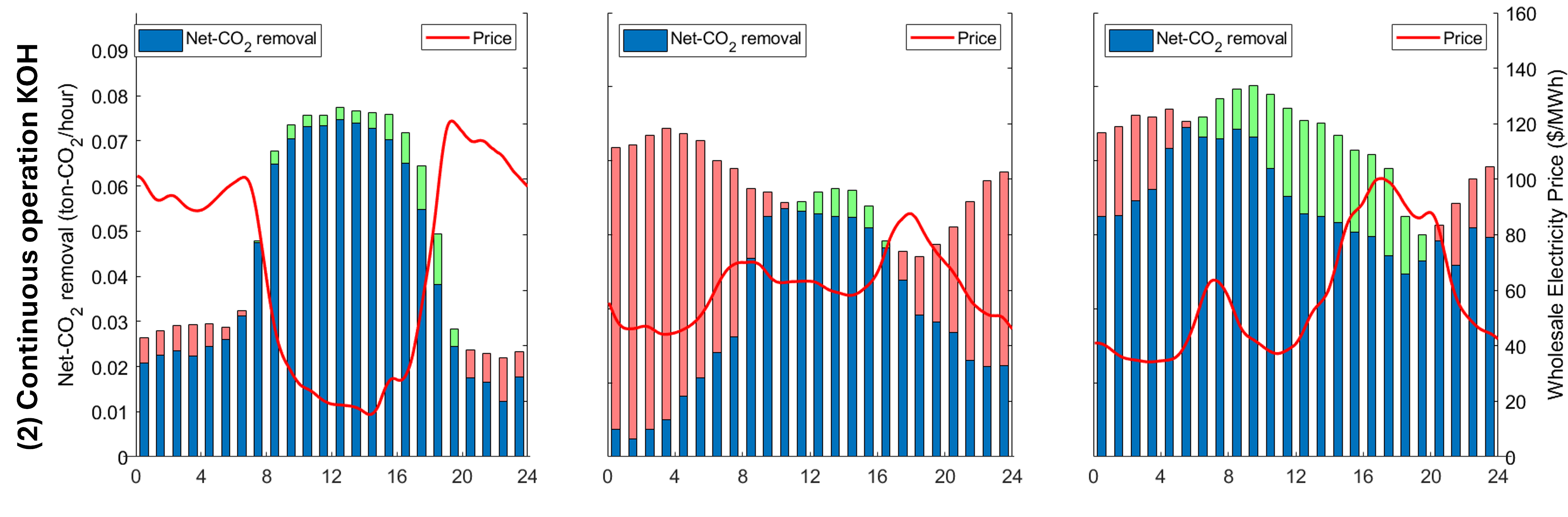} 
    \caption{\textbf{Daily wholesale electricity price profiles and DAC hourly operations.} Columns: (a) CA; (b) NY; (c) TX. Rows: (1) MOF; (2) KOH. \textcolor{red}{Red} and \textcolor{green}{green} bar represents \co abatement \textcolor{red}{loss} and \textcolor{green}{gain} due to ambient temperature and relative humidity change compared to baseline. Incentive selling price = \$200/ton-\co for MOF and \$100/ton-\co for KOH, consistent with each technology's techno-economic analysis cost result, making them marginally profitable using average electricity price. Averaged across all days in the full-year period.}
    \label{tab:fig4}
\end{figure*}

A more detailed look at the annual temporal behavior in different locations explains the economic-climate trade-offs. By using the MOF and KOH technologies as an example, the distributions of net-\co removal (\textcolor{blue}{Figure~\ref{tab:fig2}}) is negatively correlated with electricity price across different markets, and consistent across DAC technologies.
Results show a strong seasonal pattern. NY performs best during spring and fall when prices are low due to low heating and cooling demands~\cite{nyiso_new_2023}; CA has its worst period in November and December due to the price surge caused by unexpected cold waves~\cite{caiso_california_2023}; while TX shows summer is the worst season due to its high cooling demand and low wind profile~\cite{ercot_electric_2023}. The seasonal price pattern is driven by supply-demand balance of electricity, fundamentally determined by climate patterns such as temperature and relative humidity. The impact of climate on DAC technology selection is non-trivial: (1) different DAC technologies have different preferred climate conditions; (2) the climate impact can be both positive and negative, causing gain or loss in abatement capacity. 

\noindent
\begin{minipage}[t]{\columnwidth}
  \raggedright
  \includegraphics[width=\columnwidth]{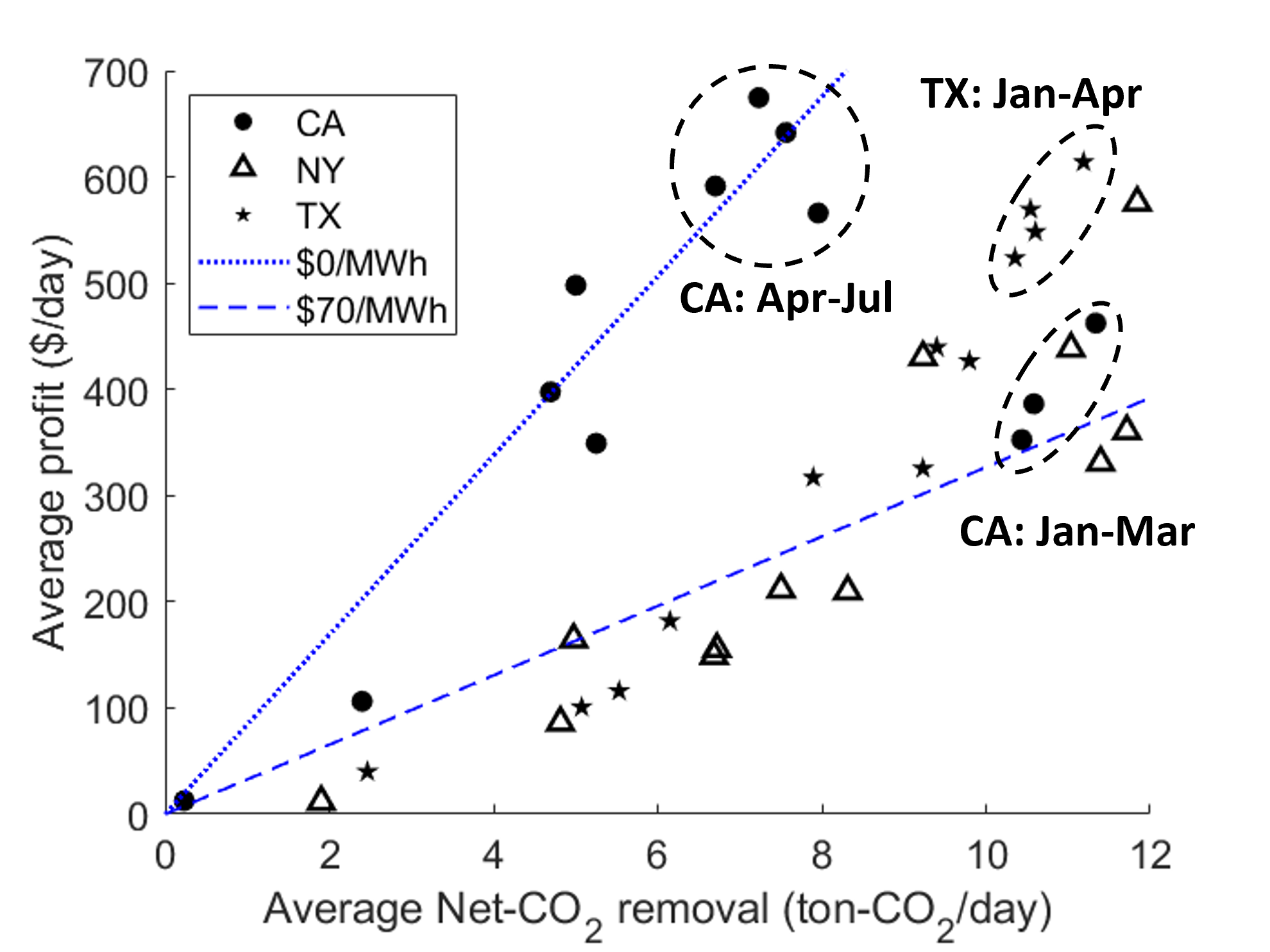}
  \captionof{figure}{\textbf{MOF DAC monthly economic–climate trade-off for profit-driven DAC operation.}
  Incentive selling price = \$200 per ton-CO$_2$.}
  \label{fig3}
\end{minipage}

\vspace{6pt}

 Cumulatively, CA surpasses TX as the most profitable location for DAC deployment, followed by NY, despite its highest average electricity price and lower net-\co removal. A strong positive correlation between monthly profit and \co removal is observed in NY and TX which grow proportionally by using \$50-70/MWh fixed electricity price (\textcolor{blue}{Figure~\ref{fig3}}). However, the most profitable season in CA (Apr-Jul) does not provide the highest \co removal, and it uses essentially free electricity. For Jan-Mar, profits and \co removals in CA are similar to NY.  In summary, local demands such as heating/cooling and load-following power services strongly affect the power system pricing and resilience. These factors need careful consideration if DAC systems are tob be deployed at a large scale.

\subsection{Daily Temporal Behaviors}\label{subsec2}

We examine the hourly resolution profile from daily average price (aggregated from 5-min resolution) for DAC operations and sheds light on the divergence between monthly profits and \co removal in different states. CA shows a distinct ``duck-curve'' daily pattern that prices are significantly lower during the day~(\textcolor{blue}{Figure~\ref{tab:fig4}}), while in NY and TX, the wholesale electricity prices are relatively stable within a day, showing a typical 2-peak profile \cite{ruggles_developing_2020}. As a result, DAC systems in CA would operate at a high capacity factor during the daytime while remaining largely idle in other periods. Notably, the daytime price in CA is often near zero or even negative during spring months (Apr-Jul) when the cooling demand is also low, yielding large profit potential despite the net \co removal being lower than in NY or TX. This example shows how DAC operations could benefit from additional renewable deployments which increase grid volatility but contribute to more low-price intervals. Regional climate conditions cause significant capture reduction for NY, while TX favoring KOH than MOF. DAC operations are relatively insensitive to CA's climate conditions for both technologies. Different DAC technologies exhibit varying sensitivities to ambient temperature and humidity. Sorbent-based systems (e.g., MOF) perform best in temperate, low-humidity environments, as water vapor interferes with \co adsorption and high temperatures weaken binding strength. In contrast, KOH-based liquid solvent systems benefit from hot and humid climates, where water availability and elevated temperatures enhance \co absorption efficiency by forming Ca(OH)\textsubscript{2}. \rev{Ambient temperature–humidity corrections generally act as efficiency adjustments, affecting about 5–20\% of net \co removal. However, when conditions reach tipping points (e.g., KOH technology in NY during nighttime), they can flip profitable operations into losses, leading to large uncertainty in optimization outcomes.}


\subsection{Importance of Temporal Resolution}

\textcolor{blue}{Figure~\ref{tab:fig4}(c)} shows that aggregating daily results to monthly averages can mask important details. For TX, July's average price is inflated by some expensive days, even though 8 days had prices below \$60/MWh, keeping DAC marginally profitable for operation. In contrast, all 31 days in August had prices above \$60/MWh but don't have any extreme high price outliers, making DAC operations largely unprofitable for all days and reducing its capacity factor to near 0 (see \textcolor{blue}{Figure S3}). This mismatch underscores that high temporal resolution analysis is essential for accurately capturing power market volatility and guiding DAC operations.

Flexible DAC operations for profit maximization is rational for investments compared to full-capacity operations under grid-average properties when incentives are not sufficiently high. It can significantly improve profitability, turning some otherwise unprofitable projects into viable ones and accelerating investment payback periods—thereby encouraging new DAC deployments. It is also necessary to avoid possible misleading differences from using grid-average assumptions. Marginal pricing in power markets automatically synchronizes DAC operation with renewable profiles. The simultaneity can boost profit and carbon removal efficiency, absorbing additional/curtailed renewable, stabilizing power prices, but limit the DAC operation with renewable capacity factor. This approach may extend to other flexible demands, such as green hydrogen production \cite{ruhnau_flexible_2023}, and could motivate mobile DAC systems, like rail-based solutions leveraging different low-price periods in different markets\cite{bachman_rail-based_2022}. Analyzing power market temporal behavior further helps optimize maintenance and labor, reducing costs.

\section{Policy Implications}\label{sec4}

\begin{figure*}[h]
    \centering
    \includegraphics[width=0.75\linewidth]{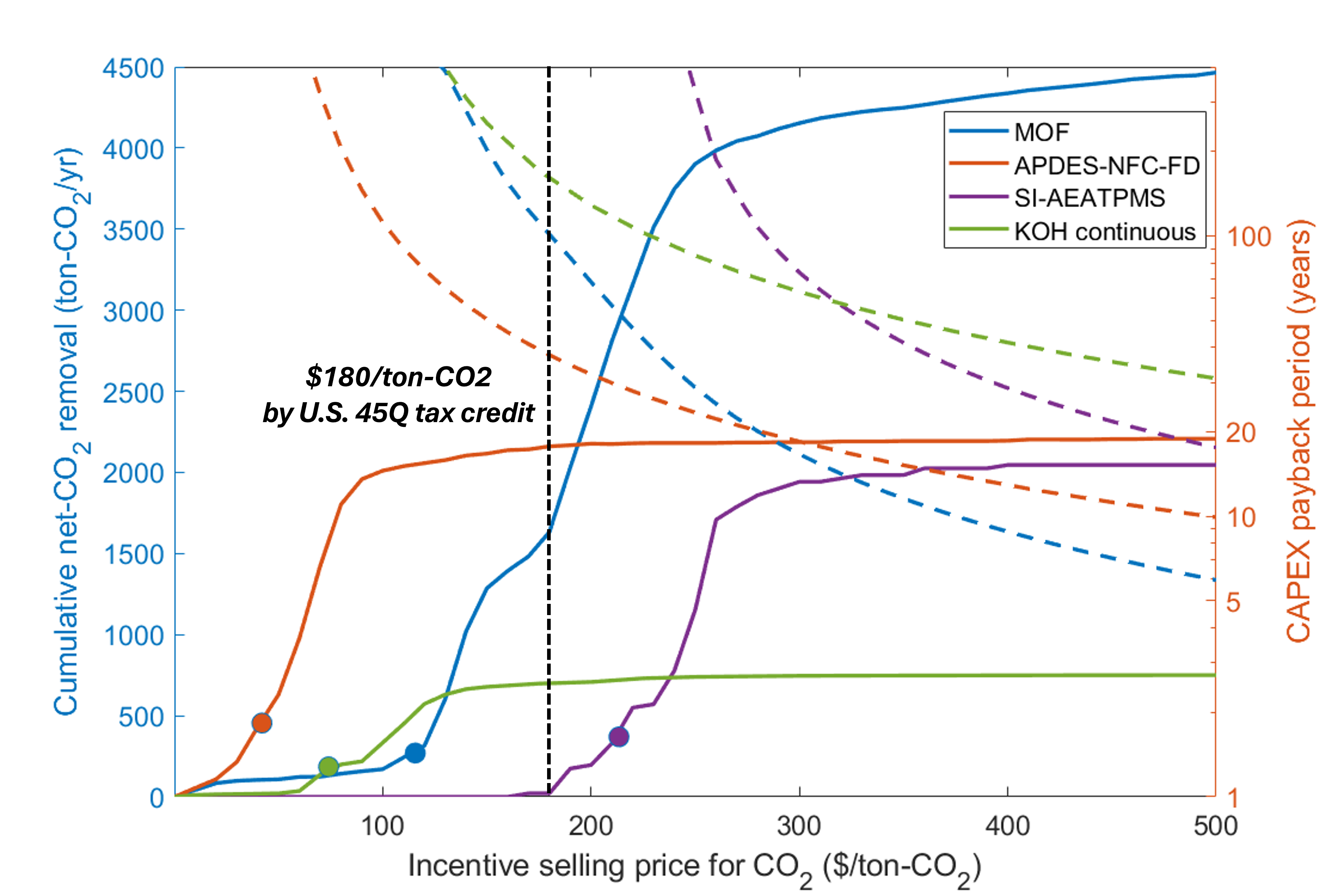}
    \caption{\textbf{Annual cumulative net-\co removal (left, solid) and CAPEX payback years (right, dashed) for different DAC technologies with the same investment in CA power market.} The vertical dashed line highlights the current U.S. 45Q legislation \$180/ton tax credit incentive for DAC geological storage. Anchoring points for each technology pinpoints where the incentive equals the cycle cost (material cost + thermal cost) for each DAC technology respectively. Right y-axis showing impact of CAPEX and capital payback years. }
    \label{tab:fig5}
\end{figure*}

While DAC profit maximization is rational from an economic perspective, it's essential to recognize that \co differs significantly from conventional commodities. Unlike most product economics driven by supply and demand, the value of \co is primarily rooted in the urgent need to mitigate its adverse impact on global warming \cite{groom_social_2023}. Although commercial opportunities for \co utilization such as cement \cite{shah_cement_2022}, synthetic fuels and chemicals \cite{bhardwaj_opportunities_2021} are growing, the need to scale \co removal have been shaped by government regulations through national policies, rather than market forces. For DAC operations, the profit arises only after the retirement of credits in a voluntary or compliance \co market. Therefore, a careful policy design by optimizing the incentives for DAC operations or imposing carbon taxes on their energy consumption, is pivotal. Better incentive policy design can address the economic-climate trade-off. However, applying carbon taxes to energy consumption for DAC plants may be counterproductive.

\subsection{\co Incentives}\label{subsec3}

Using the CA as an example, we evaluate the four DAC technologies under different \co incentive policies, i.e., revenues received by selling captured \co credits (\textcolor{blue}{Figure~\ref{tab:fig5}}). Although very different in assumptions, all DAC technologies exhibit a qualitatively consistent trend. 

All DAC technologies are assumed to have the same initial investments of total \$10.5M, but different CAPEX assumptions for each different plant (see \textcolor{blue}{Table}~\ref{tab:ben}). MOF technology plant delivers the highest net-\co removal if the incentive is sufficiently high. However, its relatively high OPEX makes it less competitive than AN technology when incentives are low. Both exceed the net \co removals using SA and KOH technologies which have high CAPEX or OPEX. The net-\co removal grows quickly beyond the anchoring points, where the incentive equals the non-electricity OPEX, referred as ``cycle cost'' for sorbent DAC systems. With electricity price volatility, the cycle cost sets a lower threshold for initiating \co removal, while CAPEX defines the upper limit for the total \co removal capacity.

Implementing DAC operation to maximize profit is not a simple binary decision. The amount of total \co removal and total profit grow gradually with the increasing \co incentives. \co removal plateaus when it approaches a DAC plant's maximum capacity, but profit keeps growing as the incentive increase, leading to shorter capital payback years. When incentive is sufficiently large, low CAPEX technology pays back earlier even with lower energy efficiency. Net \co removal volume is quite sensitive to the range of incentives applied where a small change can impact \co removal significantly. In the U.S. for instance, the current 45Q legislation \cite{us_congress_inflation_2022} provides a \$180/ton tax credit for DAC with geological storage. This falls within the sensitive price range for MOF technology (\textcolor{blue}{Figure~\ref{tab:fig5}}), suggesting a large opportunity cost: increasing the 45Q incentive from \$180/ton to \$250/ton would mean that using MOF is economically more favorable than using AN, and approximately doubles the net \co removal volume.

\begin{figure*}[h!]
    \centering
            \includegraphics[width=0.49\textwidth]{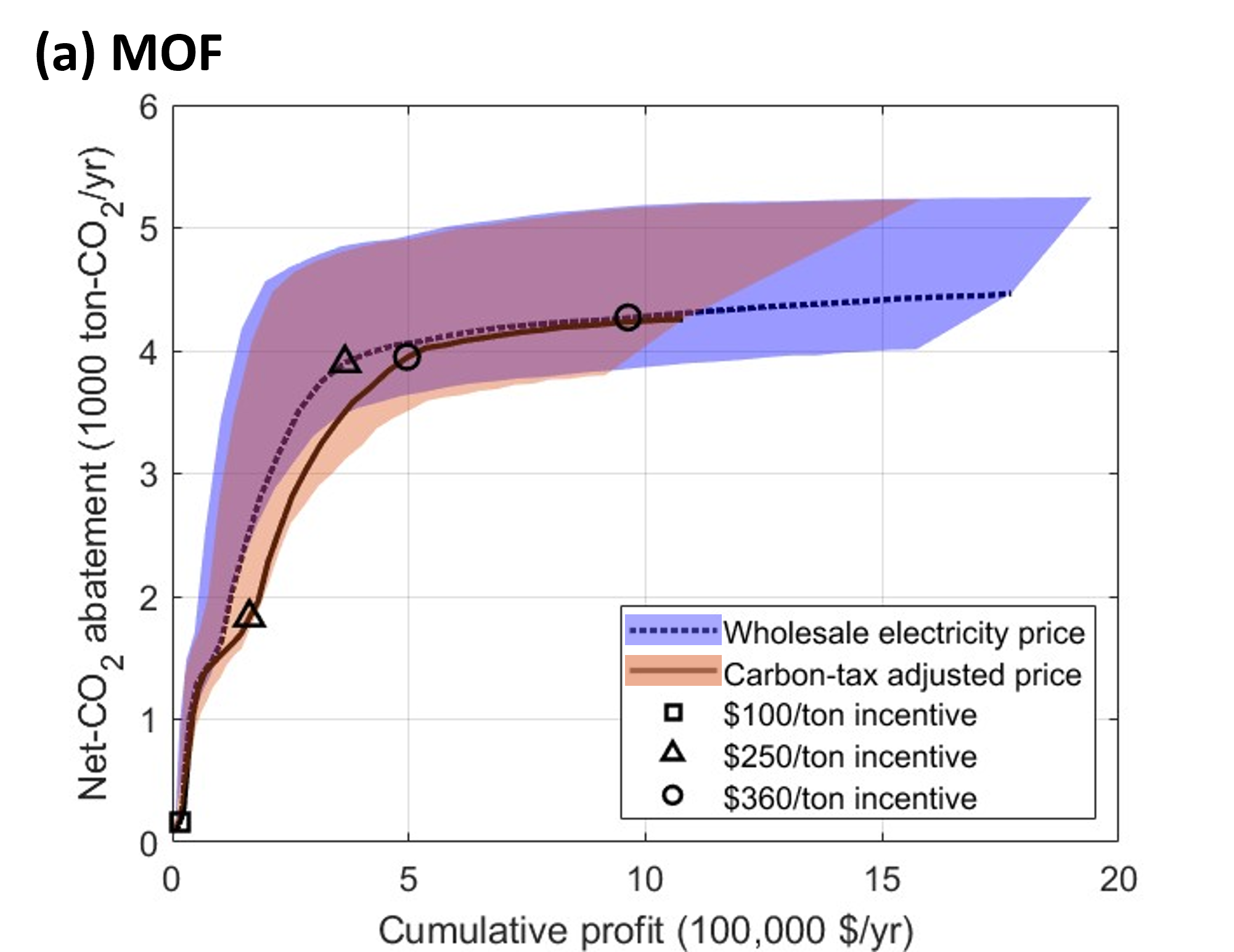}
            \includegraphics[width=0.49\textwidth]{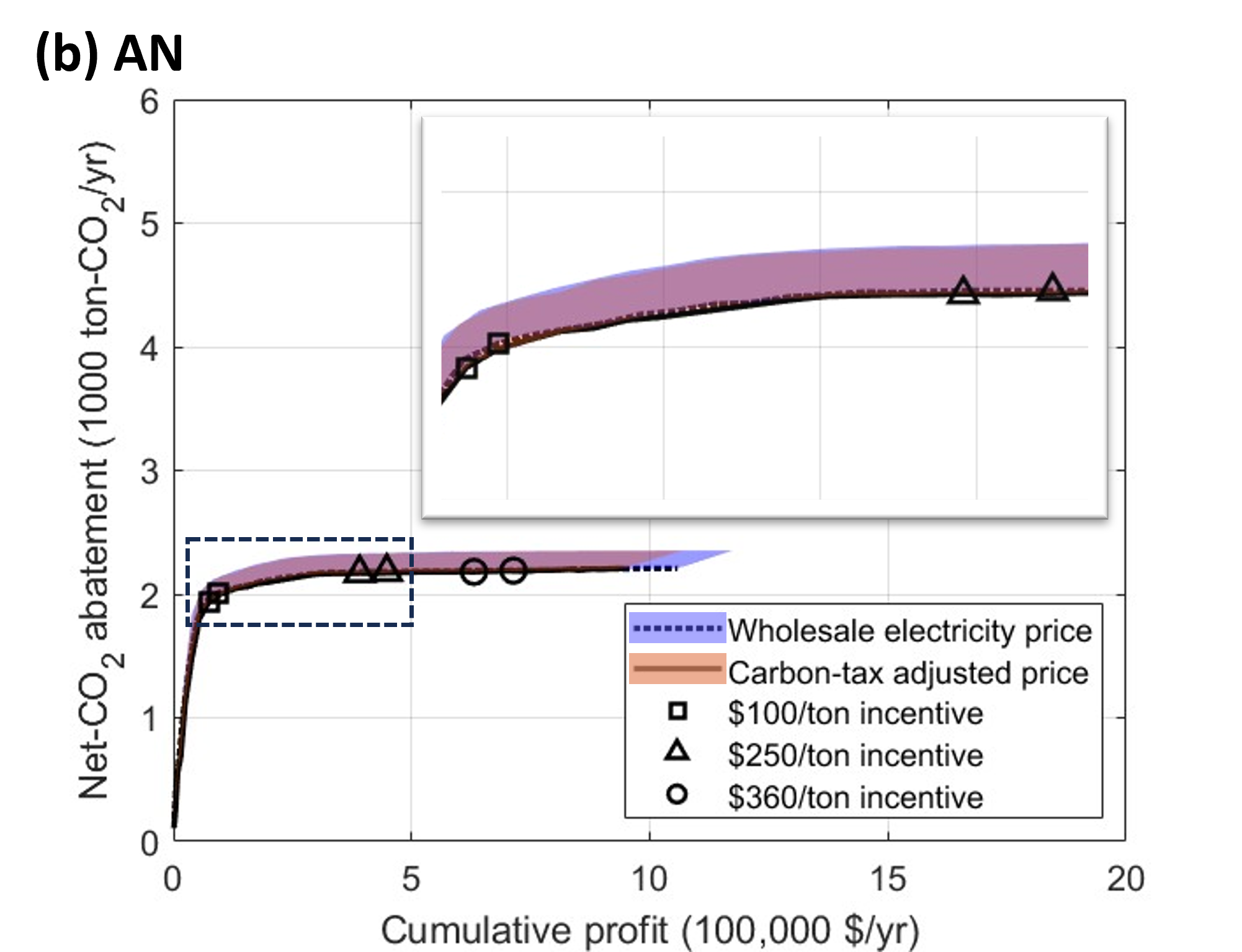} 
            \includegraphics[width=0.49\textwidth]{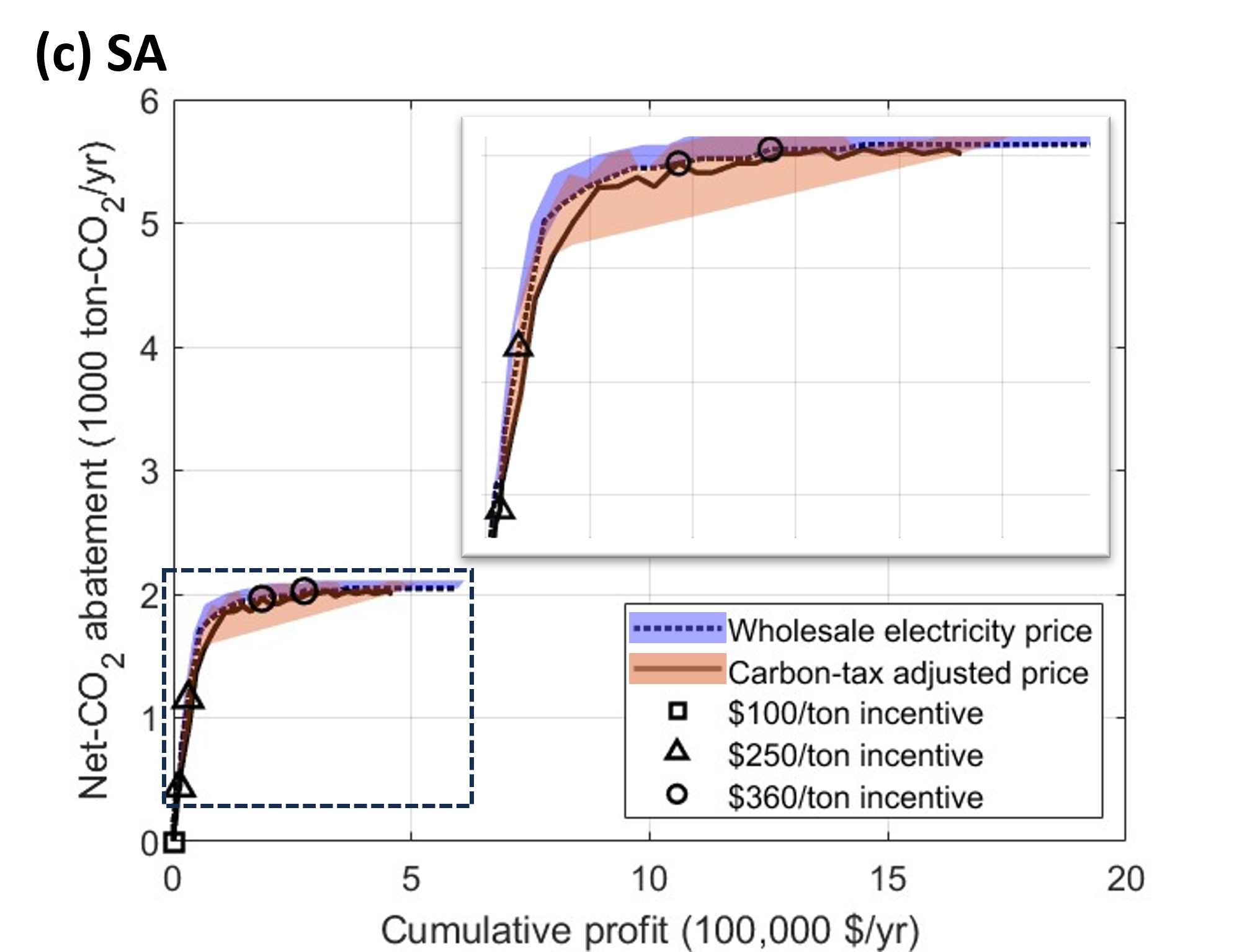} 
            \includegraphics[width=0.49\textwidth]{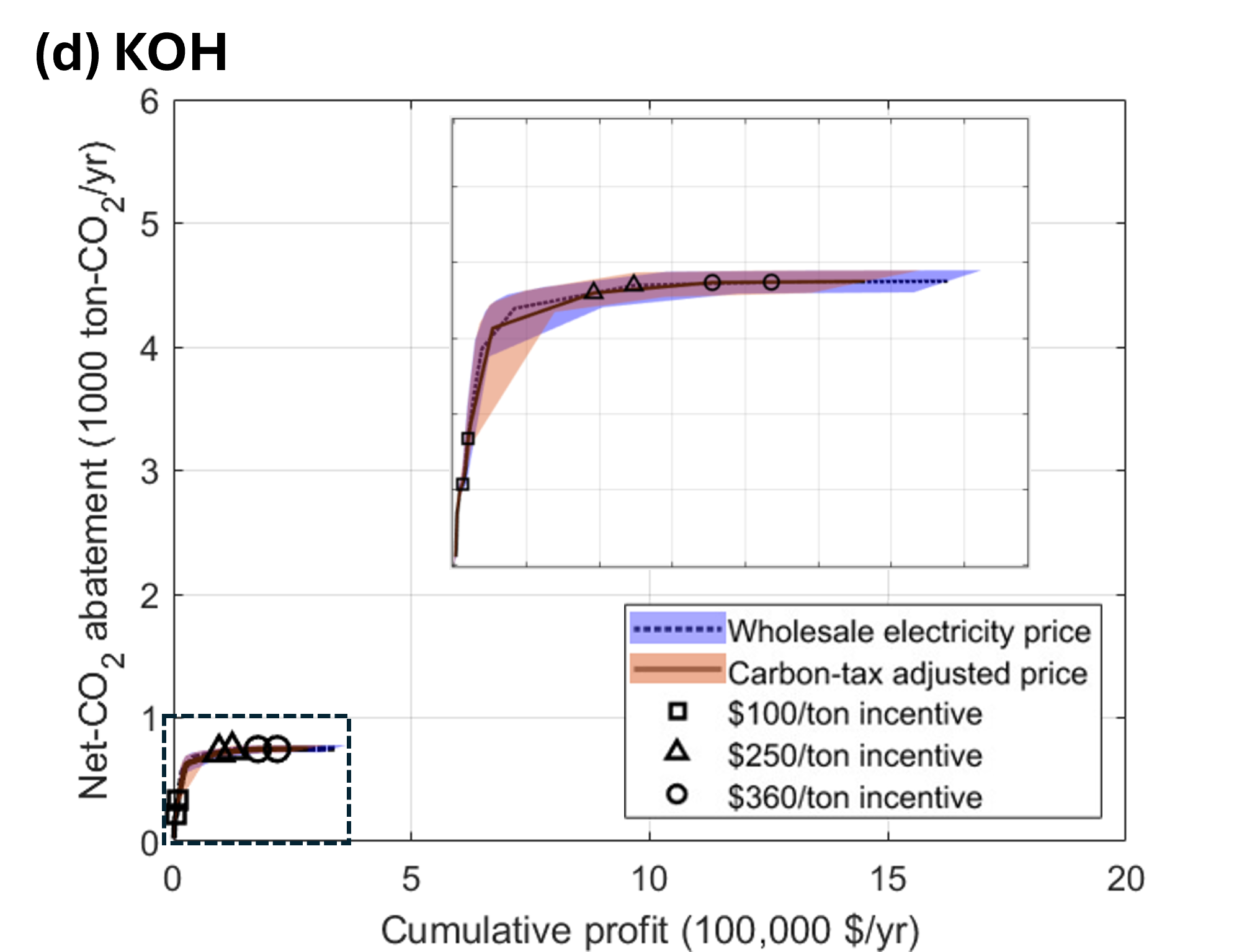} 
    \caption{\textbf{Profit-removal relationship and carbon-tax impact in CA power market} (a) MOF; (b) AN; (c) SA; (d) KOH. The shaded area represents the region covered by multiple sensitivity analyses. Figs (b),(c), and (d) provide a zoom-in detailed comparison. Three anchoring points (squares, triangles, circles) indicate the optimal incentives policy at corner points for MOF and AN technologies. Optimal incentives for MOF are \$250/ton-\co using wholesale electricity price, and \$360/ton-\co using carbon-tax adjusted electricity price. For AN, optimal incentives are \$100/ton-\co for both with/without carbon tax cases with negligible differences between two price cases. All prices assume the carbon-tax applied to electricity equals the incentive offered.}
    \label{tab:fig6}
\end{figure*}

\textcolor{blue}{Figure~\ref{tab:fig6}} compares \co removal and profit for each DAC technology under different incentive assumptions. \co removal initially increases with greater incentives for all DAC systems. Beyond a sharp corner point, \co removal plateaus and additional incentive value translates into additional profit. This corner point corresponds to the optimal incentive policy design for the \co ``selling price" and differs for each DAC technology. Using this information as a tool will be helpful in designing an improved DAC incentive policy and optimizing the economic-climate trade-offs, as well as tracking ``the state-of-the-art technology''.   

\subsection{Carbon-tax}\label{subsec3}

With point-source carbon capture and storage systems (CCS), such as power plants, applying carbon tax on energy consumption may be a positive driving factor that encourages deployment \cite{sun_impact_2022, victor_ccus_2022}.  Emission baselines are positive and will be reduced for such systems, lowering commerial carbon tax expenses and making CCS an economically rational decision. Therefore, a carbon tax applied in these cases can lead to net \co reductions. For DAC \co removal, however, low-emission electricity consumption driven by a carbon tax optimizes capture efficiency while sacrificing capacity factor, and result in effectively higher electricity costs. We therefore consider an analysis of the effect of a carbon tax applied to electricity consumption in DAC operations. \rev{Since a carbon tax alters the electricity price signal but does not eliminate associated emissions, these emissions are still accounted for in incentive payments, thereby shifting the optimization outcome; the coexistence of both policies may reflect different valuation of \co (in \$/ton) in practice and does not introduce double counting or conflict.} 

Results show that the trade-off between capture efficiency and net \co removal volume for DAC technoloiges is highly skewed, strongly favoring higher removal volume when costs (including taxes) are reduced. The imposition of a carbon tax on electricity use may harm both the net \co removal and the profitability of potential DAC operations. \textcolor{blue}{Figure~\ref{tab:fig6}} shows that a carbon tax always falls within or below the raw wholesale price case for \co removals for all DAC technologies. While the carbon tax marginally enhances \co capture efficiency, it fails to compensate for the substantial reduction in net \co removal volume. For the MOF technology, which has high electricity consumption, applying a carbon tax significantly increases the needed to reach the incentive the same optimal corner point (i.e. \$250/ton increases to \$360/ton). For AN technology, with lower energy requirements, a carbon tax still generates lower profit and no improvement in net \co removal volume.  The added cost from the carbon tax outweighs the benefit of increased capture efficiency, rendering the system unprofitable to stop operation and resulting in less \co removal instead of efficiency gain. The impact of a carbon tax on energy use for DAC operations is therefore negative, irrespective of the technology used or the \co removal incentive offered. This means that a carbon tax on electricity use may be counterproductive for both profitability and net \co removal of DAC plant operations.

\begin{figure*}[b]
        \centering
            \includegraphics[width=0.49\textwidth]{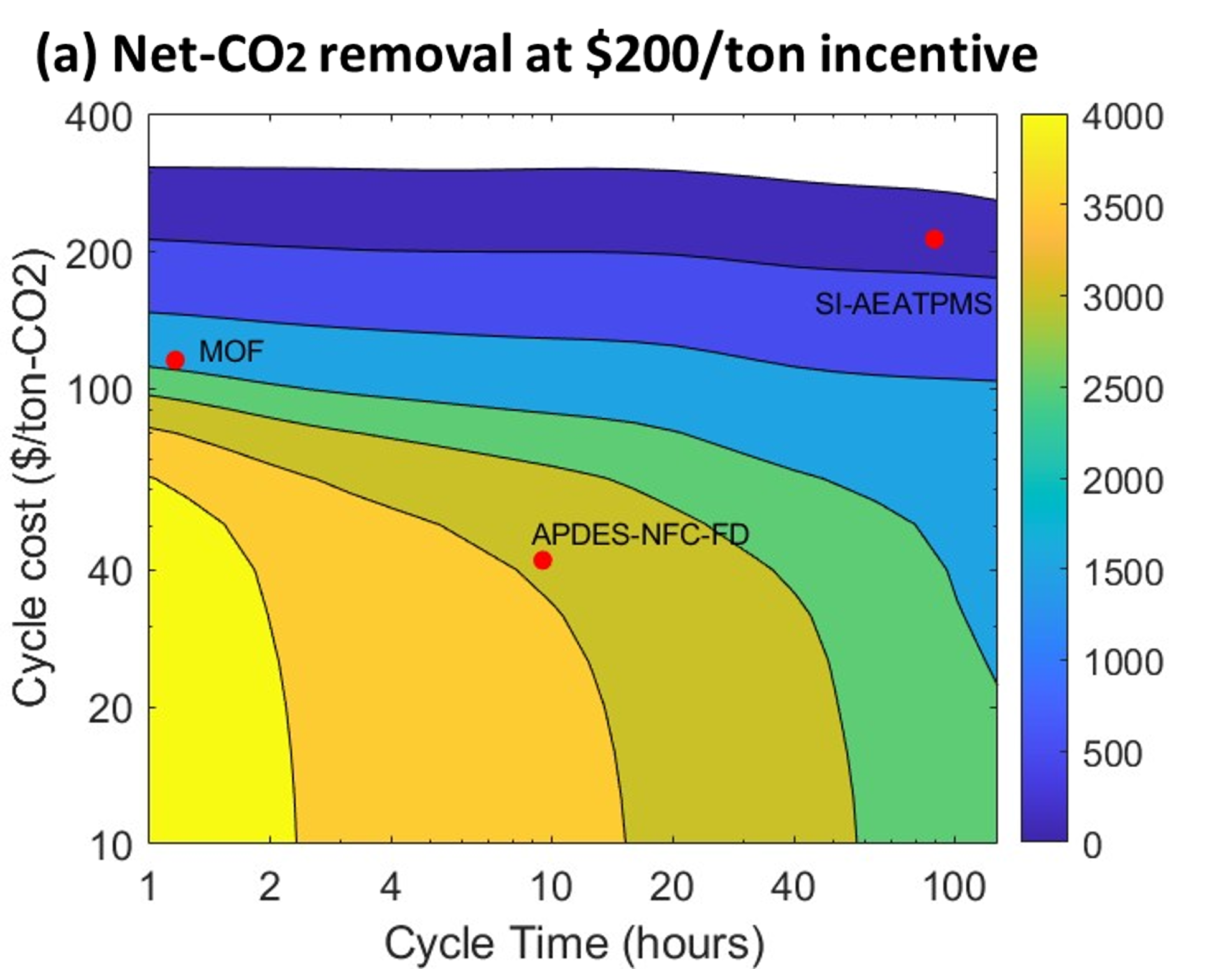}
            \includegraphics[width=0.49\textwidth]{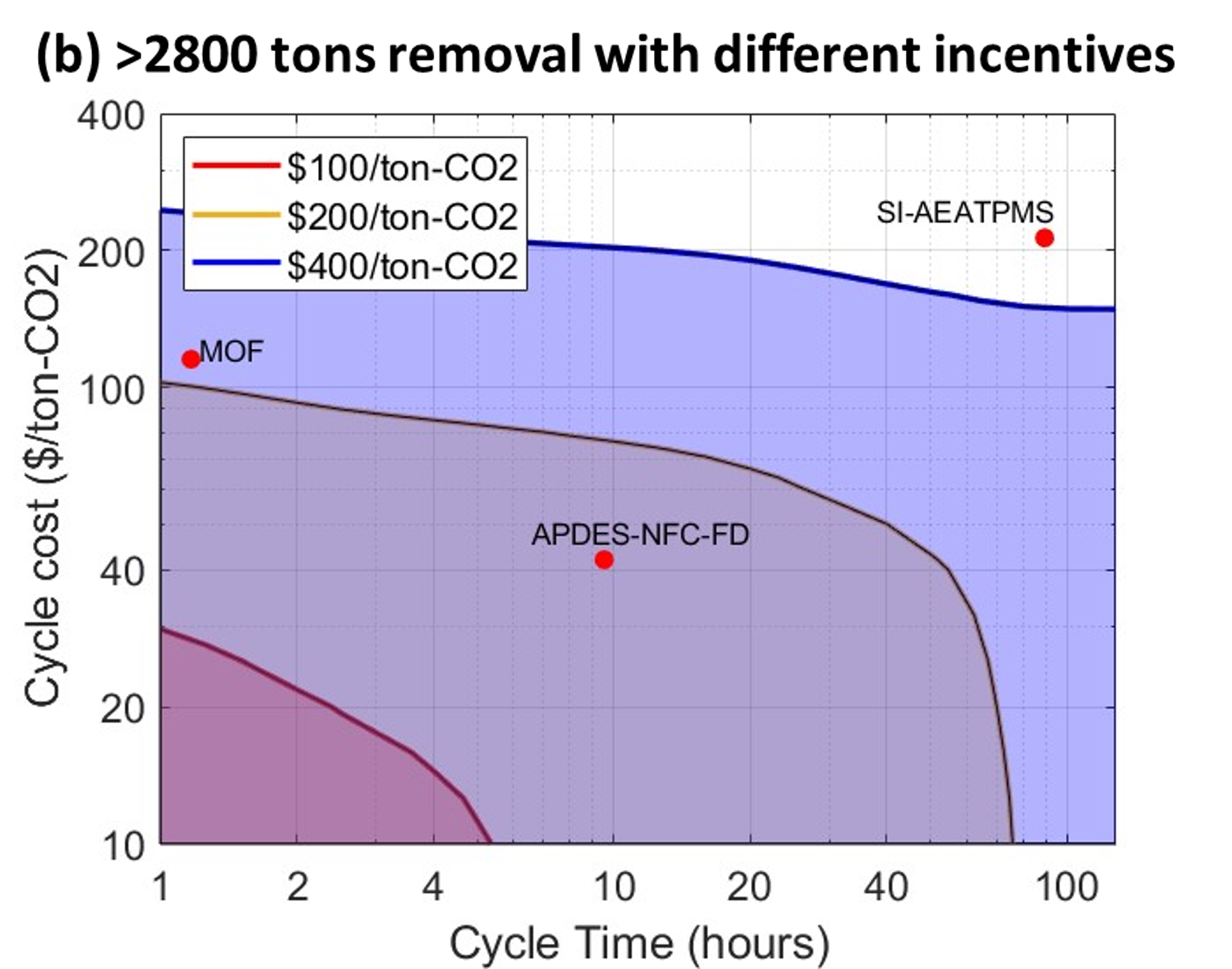} 
        \caption{\textbf{DAC design space with cycle costs and cycle time in CA power market. (a)} Net-\co removal with \$200/ton incentive. \textbf{(b)} DAC designs that deliver \(>\) 2800 ton annual removal with different incentives with overlapping coverages. Different cycle costs (y-axis) tests the sensitivity to the two components of "cycle costs": (1) sorbent materials cost; (2) thermal energy cost. Results are smoothed with Gaussian smoothing kernel of standard deviation of 1.5 to account for uncertainties in the algorithm. \rev{This design space analysis illustrates how varying the cycle cost (containing both material and thermal energy costs) shifts the optimization outcome, serving as a sensitivity analysis for both thermal energy cost and sorbent material cost.}}
        \label{tab:fig7}
\end{figure*}

Eventually, a carbon tax policy on DAC electricity usage will not lead to more CO2 removal. We suggest an exemption for DAC plants from such taxes or equivalent spending (e.g., require purchasing of renewable energy certificates - RECs) on condition that their capture efficiency remains reasonable, perhaps at least 75\%-80\%, which should be  set based on specific power market and technology requirements.

\section{DAC Design Implications}\label{sec5}

Comparing different DAC technologies (\textcolor{blue}{Figure~\ref{tab:fig6}} and \textcolor{blue}{Supplementary Information} for more details) reveals that reducing system CAPEX, material costs, and cycle time has a significant impact on DAC operations.

Profit-driven operations are very sensitive to CAPEX and material costs, offering a substantial room for improvement. DAC adsorbents must be durable, withstanding repeated regeneration cycles while retaining the effective capacity over their material lifespans \cite{low_analytical_2023}. Orders-of-magnitude differences in costs per ton removal \cite{leonzio_environmental_2022} and orders-of-magnitude differences uncertainty about material resilience \cite{sinha_systems_2017} exist for each. Additionally, CAPEX estimation is highly sensitive to plant scale and the technology chosen, and thus technical improvements are reported to affect performance by a factor of 2 to 10 \cite{fasihi_techno-economic_2019}.

Shorter cycle times and faster adsorption and desorption rates also improve DAC performance. Accurate cycle time assumptions are often missing in techno-economic analysis for full-capacity operations, and when high-resolution temporal volatility is introduced, long-cycle operations may no longer remain profitable (\textcolor{blue}{Figure~\ref{tab:fig7}}). Flexibility in cycle times may be critical as price incentives change in different power markets. From our market integration analysis, an ideal target for DAC development may have cycle periods of 1 to 2 hours and cycle costs below \$80/ton of \co captured (not accounting for electricity cost). With new research and development in this area, routine day-ahead power markets and hourly-resolution dispatch prices, DAC operations may eventually be able to work with hourly market clearing resolution with short cycle time. In summary, reducing system CAPEX, material costs, and cycle times has the potential to achieve the cost efficiencies necessary to ultimately accelerate DAC deployment in commercial markets \cite{lackner_buying_2021}.

\section{Discussion}\label{sec4}

This paper considers temperature-pressure-swing adsorption/desorption solid sorbent and KOH liquid solvent for a total of four commercial DAC technologies. Our methodology is adaptable to other cyclic or continuous DAC technologies as well, including pressure-swing, moisture-swing approaches, as well as membrane based DAC systems that operates continuously. The economic model accommodates diverse heat/electrical assumptions with varying cycle times and operational flexibility, facilitating both profit-driven operations and techno-economic analysis. It can also be adapted to simulate heat electrification by aggregating thermal energy consumptions into the desorption/regeneration phase. Despite the significant differences in the DAC technology parameters tested, our results exhibit consistent and robust qualitative trends that are broadly application across larger contexts.  

Our analysis demonstrates that the integration of DAC operations with power market volatility prompts new perspectives. For DAC system development and deployment, key implications are: (1) DAC technology development should prioritize low CAPEX (<100\$/ton-\cd-year), low material costs (<\$80/ton-\cd), and high flexibility with shorter cycle time (1-2 hours); (2) DAC should connect to appropriate power markets with preferred environmental condition such as temperature and humidity, such as liquid solvent DAC system for Texas. Profit driven operations can lead to earlier DAC deployment with higher economic efficiency, for example, MOF technology can increase profit >10 times and shorten the investment payback period by 15 years. Optimal DAC designs should be tailored to specific climates and geographic areas, where siting of DAC plants is fundamentally dictated by local climate features, both the renewable profiles in the power market and the appropriate climate conditions (e.g., temperature and humidity). Different power markets favor different DAC technologies, and should be carefully considered in decision-making for plant design and deployment. \rev{While this study applied ISO-based average emission signal for optimization, recent work~\cite{sukprasert_implications_2024} shows that average and marginal signals can diverge substantially, with the marginal signal may provide a more accurate measure of incremental load impacts. Exploring this distinction represents an important avenue for future analysis.}

For policymakers, consideration of these factors for profit-driven DAC operations may help to  establish effective policies that maximize net \co removal volume and assure that ``efficiency'' does not take precedence over ``scale''.  High \co capture efficiency and energy efficiency at the expense of plant capacity, material lifespan, operational flexibility, and economic viability are counterproductive. Good policy design prioritizes incentives that balance the economic–climate trade-offs, which can potentially double the MOF DAC removal if incentive is raised from \$180/ton to \$250/ton. Carbon tax exemptions or establishing a minimum required capacity factor of DAC plants for eligibility of government subsidies may be needed. In regulated Cap-and-Trade markets, for example, profitability can be boosted during high \co price opportunities, lowering \co removal costs for other carbon market participants. \rev{Our findings apply to a carbon tax combined with net-removal incentives, and they generally extend to variants such as tax refunds or clean procurement requirements, provided these policies effectively changes the electricity price signals, and eventually lead to shift of profit-max optimization results}

In marginal-price-based markets, profit driven DAC operation will automatically avoid peak hours with high wholesale electricity prices and absorb additional/curtailed renewables during off-peak hours. Power system operators will benefit from such flexible DAC operations and reduce system cost, emissions, and instability with “renewable–following” DAC operations. Specifically, solar simultaneity suggests that the prices can be much lower than grid averages, enabling profitability and encouraging DAC expansion. Certain market strategies and policies might also help to increase the DAC system capacity factors and maximize \co removal, such as firm power purchase agreements.

\section{Conclusion}\label{sec7}

Our global energy system is currently undergoing a fundamental transformation towards a net-zero emission target. Both \co removal using DAC and power grid decarbonization play pivotal roles in this transition. Given that this energy transition is anticipated to be lengthy and gradual, demanding an unprecedented level of investment, it is crucial to maintain the stability of the energy system throughout the transition period and ensure economic viability. With profit-driven investments in technically mature, large-scale DAC plants, operations under power market volatility can be financially advantageous, climate beneficial, and allow for a smooth energy transition. Taking the DAC system perspective for profit driven operation will introduces an economic-climate trade-off with implications extending throughout the entire lifecycle of DAC system operations, reflecting efficient use of current technologies and future technology R\&D, power markets integration, and decarbonization policy designs. Our insights and methods open up future work for potential DAC system design and sorbent materials customization for specific power market and climate zones, and provide power market services such as demand response. These are important topics to a successful low-carbon integrated power-energy system.

\printcredits
\vspace{0.3cm}

\noindent \textbf{Declaration of competing interest}

The authors declare that they have no known competing financial interests or personal relationships that could have appeared to influence the work reported in this paper.
\vspace{0.3cm}

\noindent \textbf{Data and code availability}

The original data can be downloaded from: \cite{copernicus_climate_change_service_era5_2018} for ambient temperature and relative humidity data, and Electricity price/emission data from official websites for \href{https://www.nyiso.com/energy-market-operational-data}{NY}, \href{http://oasis.caiso.com/mrioasis/logon.do}{CA}, and \href{https://www.ercot.com/mp/data-products/data-product-details?id=NP6-785-ER}{TX}. All model and data used can be accessed from \href{https://github.com/ZhiyuanF/DAC_optimization}{Github}. 
\vspace{0.3cm}

\noindent \textbf{Acknowledgements}

This report was partially funded through a gift from the Alfred P. Sloan Foundation and the Center on Global Energy Policy. More information is available at \url{https://energypolicy.columbia.edu/about/partners}. This report represents the research and views of the authors. It does not necessarily represent the views of the Center on Global Energy Policy. The piece may be subject to further revision.

\bibliographystyle{unsrt}

\bibliography{cas-refs-1}

\begin{thebibliography}{10}

\bibitem{lee_ipcc_2023}
Katherine Calvin, Dipak Dasgupta, Gerhard Krinner, Aditi Mukherji, Peter~W. Thorne, Christopher Trisos, and et~al.
\newblock {IPCC}, 2023: {Climate} {Change} 2023: {Synthesis} {Report}. {Contribution} of {Working} {Groups} {I}, {II} and {III} to the {Sixth} {Assessment} {Report} of the {Intergovernmental} {Panel} on {Climate} {Change} [{Core} {Writing} {Team}, {H}. {Lee} and {J}. {Romero} (eds.)]. {IPCC}, {Geneva}, {Switzerland}.
\newblock Technical report, Intergovernmental Panel on Climate Change (IPCC), July 2023.
\newblock Edition: First.

\bibitem{lamboll_assessing_2023}
Robin~D. Lamboll, Zebedee R.~J. Nicholls, Christopher~J. Smith, Jarmo~S. Kikstra, Edward Byers, and Joeri Rogelj.
\newblock Assessing the size and uncertainty of remaining carbon budgets.
\newblock {\em Nature Climate Change}, October 2023.

\bibitem{tollefson2023too}
Jeff Tollefson.
\newblock Is it too late to keep global warming below 1.5° c? the challenge in 7 charts.
\newblock {\em Nature}, 2023.

\bibitem{buck_why_2023}
Holly~Jean Buck, Wim Carton, Jens~Friis Lund, and Nils Markusson.
\newblock Why residual emissions matter right now.
\newblock {\em Nature Climate Change}, 13(4):351--358, April 2023.

\bibitem{marcucci_road_2017}
Adriana Marcucci, Socrates Kypreos, and Evangelos Panos.
\newblock The road to achieving the long-term {Paris} targets: energy transition and the role of direct air capture.
\newblock {\em Climatic Change}, 144(2):181--193, September 2017.

\bibitem{goldberg_co-location_2013}
David~S. Goldberg, Klaus~S. Lackner, Patrick Han, Angela~L. Slagle, and Tao Wang.
\newblock Co-{Location} of {Air} {Capture}, {Subseafloor} {CO} $_{\textrm{2}}$ {Sequestration}, and {Energy} {Production} on the {Kerguelen} {Plateau}.
\newblock {\em Environmental Science \& Technology}, 47(13):7521--7529, July 2013.

\bibitem{mcqueen_review_2021}
Noah McQueen, Katherine~Vaz Gomes, Colin McCormick, Katherine Blumanthal, Maxwell Pisciotta, and Jennifer Wilcox.
\newblock A review of direct air capture ({DAC}): scaling up commercial technologies and innovating for the future.
\newblock {\em Progress in Energy}, 3(3):032001, July 2021.

\bibitem{alcalde_estimating_2018}
Juan Alcalde, Stephanie Flude, Mark Wilkinson, Gareth Johnson, Katriona Edlmann, Clare~E. Bond, Vivian Scott, Stuart M.~V. Gilfillan, Xènia Ogaya, and R.~Stuart Haszeldine.
\newblock Estimating geological {CO2} storage security to deliver on climate mitigation.
\newblock {\em Nature Communications}, 9(1):2201, June 2018.

\bibitem{parigi_power--fuels_2019}
Davide Parigi, Emanuele Giglio, Alicia Soto, and Massimo Santarelli.
\newblock Power-to-fuels through carbon dioxide {Re}-{Utilization} and high-temperature electrolysis: {A} technical and economical comparison between synthetic methanol and methane.
\newblock {\em Journal of Cleaner Production}, 226:679--691, July 2019.

\bibitem{gulzar_carbon_2020}
Arif Gulzar, Aanisa Gulzar, Mohd~Bismillah Ansari, Fei He, Shili Gai, and Piaoping Yang.
\newblock Carbon dioxide utilization: {A} paradigm shift with {CO2} economy.
\newblock {\em Chemical Engineering Journal Advances}, 3:100013, November 2020.

\bibitem{liu_new_2021}
Baoju Liu, Jiali Qin, Jinyan Shi, Junyi Jiang, Xiang Wu, and Zhihai He.
\newblock New perspectives on utilization of {CO} 2 sequestration technologies in cement-based materials.
\newblock {\em Construction and Building Materials}, 272:121660, February 2021.

\bibitem{fasihi_techno-economic_2019}
Mahdi Fasihi, Olga Efimova, and Christian Breyer.
\newblock Techno-economic assessment of {CO2} direct air capture plants.
\newblock {\em Journal of Cleaner Production}, 224:957--980, July 2019.

\bibitem{sabatino_comparative_2021}
Francesco Sabatino, Alexa Grimm, Fausto Gallucci, Martin Van Sint~Annaland, Gert~Jan Kramer, and Matteo Gazzani.
\newblock A comparative energy and costs assessment and optimization for direct air capture technologies.
\newblock {\em Joule}, 5(8):2047--2076, August 2021.

\bibitem{wiegner_optimal_2022}
Jan~F. Wiegner, Alexa Grimm, Lukas Weimann, and Matteo Gazzani.
\newblock Optimal {Design} and {Operation} of {Solid} {Sorbent} {Direct} {Air} {Capture} {Processes} at {Varying} {Ambient} {Conditions}.
\newblock {\em Industrial \& Engineering Chemistry Research}, 61(34):12649--12667, August 2022.

\bibitem{jiang_sorption_2023}
L.~Jiang, W.~Liu, R.Q. Wang, A.~Gonzalez-Diaz, M.F. Rojas-Michaga, S.~Michailos, M.~Pourkashanian, X.J. Zhang, and C.~Font-Palma.
\newblock Sorption direct air capture with {CO2} utilization.
\newblock {\em Progress in Energy and Combustion Science}, 95:101069, March 2023.

\bibitem{ozkan_current_2022}
Mihrimah Ozkan, Saswat~Priyadarshi Nayak, Anthony~D. Ruiz, and Wenmei Jiang.
\newblock Current status and pillars of direct air capture technologies.
\newblock {\em iScience}, 25(4):103990, April 2022.

\bibitem{smith_biophysical_2016}
Pete Smith, Steven~J. Davis, Felix Creutzig, Sabine Fuss, Jan Minx, Benoit Gabrielle, Etsushi Kato, Robert~B. Jackson, Annette Cowie, Elmar Kriegler, Detlef~P. Van~Vuuren, Joeri Rogelj, Philippe Ciais, Jennifer Milne, Josep~G. Canadell, David McCollum, Glen Peters, Robbie Andrew, Volker Krey, Gyami Shrestha, Pierre Friedlingstein, Thomas Gasser, Arnulf Grübler, Wolfgang~K. Heidug, Matthias Jonas, Chris~D. Jones, Florian Kraxner, Emma Littleton, Jason Lowe, José~Roberto Moreira, Nebojsa Nakicenovic, Michael Obersteiner, Anand Patwardhan, Mathis Rogner, Ed~Rubin, Ayyoob Sharifi, Asbjørn Torvanger, Yoshiki Yamagata, Jae Edmonds, and Cho Yongsung.
\newblock Biophysical and economic limits to negative {CO2} emissions.
\newblock {\em Nature Climate Change}, 6(1):42--50, January 2016.

\bibitem{bose_challenges_2024}
Saptasree Bose, Debabrata Sengupta, Thomas~M. Rayder, Xiaoliang Wang, Kent~O. Kirlikovali, Ali~K. Sekizkardes, Timur Islamoglu, and Omar~K. Farha.
\newblock Challenges and {Opportunities}: {Metal}–{Organic} {Frameworks} for {Direct} {Air} {Capture}.
\newblock {\em Advanced Functional Materials}, 34(43):2307478, October 2024.

\bibitem{young_process-informed_2023}
John Young, Fergus Mcilwaine, Berend Smit, Susana Garcia, and Mijndert Van Der~Spek.
\newblock Process-informed adsorbent design guidelines for direct air capture.
\newblock {\em Chemical Engineering Journal}, 456:141035, January 2023.

\bibitem{beuttler_role_2019}
Christoph Beuttler, Louise Charles, and Jan Wurzbacher.
\newblock The {Role} of {Direct} {Air} {Capture} in {Mitigation} of {Anthropogenic} {Greenhouse} {Gas} {Emissions}.
\newblock {\em Frontiers in Climate}, 1:10, November 2019.

\bibitem{terlouw_life_2021}
Tom Terlouw, Karin Treyer, Christian Bauer, and Marco Mazzotti.
\newblock Life {Cycle} {Assessment} of {Direct} {Air} {Carbon} {Capture} and {Storage} with {Low}-{Carbon} {Energy} {Sources}.
\newblock {\em Environmental Science \& Technology}, 55(16):11397--11411, August 2021.

\bibitem{breyer_carbon_2020}
Christian Breyer, Mahdi Fasihi, and Arman Aghahosseini.
\newblock Carbon dioxide direct air capture for effective climate change mitigation based on renewable electricity: a new type of energy system sector coupling.
\newblock {\em Mitigation and Adaptation Strategies for Global Change}, 25(1):43--65, January 2020.

\bibitem{daniel_techno-economic_2022}
Thorin Daniel, Alice Masini, Cameron Milne, Neeka Nourshagh, Cameron Iranpour, and Jin Xuan.
\newblock Techno-economic {Analysis} of {Direct} {Air} {Carbon} {Capture} with {CO2} {Utilisation}.
\newblock {\em Carbon Capture Science \& Technology}, 2:100025, March 2022.

\bibitem{realmonte_inter-model_2019}
Giulia Realmonte, Laurent Drouet, Ajay Gambhir, James Glynn, Adam Hawkes, Alexandre~C. Köberle, and Massimo Tavoni.
\newblock An inter-model assessment of the role of direct air capture in deep mitigation pathways.
\newblock {\em Nature Communications}, 10(1):3277, July 2019.

\bibitem{arwa_impact_2025}
Erick~O. Arwa and Kristen~R. Schell.
\newblock Impact of direct air capture process flexibility and response to ambient conditions in net-zero transition of the power grid.
\newblock {\em Applied Energy}, 386:125549, May 2025.

\bibitem{iea_world_2022}
IEA.
\newblock World {Energy} {Outlook}, 2022, October 2022.

\bibitem{world_bank_databank_2023}
World Bank.
\newblock {DataBank}: {World} {Development} {Indicators}, October 2023.

\bibitem{herzog_getting_2024}
Howard Herzog, Jennifer Morris, Angelo Gurgel, and Sergey Paltsev.
\newblock Getting real about capturing carbon from the air.
\newblock {\em One Earth}, 7(9):1477--1480, September 2024.

\bibitem{deutz_life-cycle_2021}
Sarah Deutz and André Bardow.
\newblock Life-cycle assessment of an industrial direct air capture process based on temperature–vacuum swing adsorption.
\newblock {\em Nature Energy}, 6(2):203--213, February 2021.

\bibitem{carter_private_2020}
Lauren Carter.
\newblock Private investment in climate action, 2020.

\bibitem{maniatis_impact_2022}
Georgios~I. Maniatis and Nikolaos~T. Milonas.
\newblock The impact of wind and solar power generation on the level and volatility of wholesale electricity prices in {Greece}.
\newblock {\em Energy Policy}, 170:113243, November 2022.

\bibitem{calero_duck-curve_2022}
Ivan Calero, Claudio~A. Canizares, Kankar Bhattacharya, and Ross Baldick.
\newblock Duck-{Curve} {Mitigation} in {Power} {Grids} {With} {High} {Penetration} of {PV} {Generation}.
\newblock {\em IEEE Transactions on Smart Grid}, 13(1):314--329, January 2022.

\bibitem{caiso_california_2023}
CAISO.
\newblock California {Independent} {System} {Operator}., 2023.

\bibitem{nyiso_new_2023}
NYISO.
\newblock New {York} {Independent} {System} {Operator}, 2023.

\bibitem{ercot_electric_2023}
ERCOT.
\newblock Electric {Reliability} {Council} of {Texas}, 2023.

\bibitem{an_impact_2022}
Keju An, Azharuddin Farooqui, and Sean~T. McCoy.
\newblock The impact of climate on solvent-based direct air capture systems.
\newblock {\em Applied Energy}, 325:119895, November 2022.

\bibitem{sendi_geospatial_2022}
Marwan Sendi, Mai Bui, Niall Mac~Dowell, and Paul Fennell.
\newblock Geospatial analysis of regional climate impacts to accelerate cost-efficient direct air capture deployment.
\newblock {\em One Earth}, 5(10):1153--1164, October 2022.

\bibitem{azarabadi_sorbent-focused_2019}
Habib Azarabadi and Klaus~S. Lackner.
\newblock A sorbent-focused techno-economic analysis of direct air capture.
\newblock {\em Applied Energy}, 250:959--975, September 2019.

\bibitem{sinha_systems_2017}
Anshuman Sinha, Lalit~A. Darunte, Christopher~W. Jones, Matthew~J. Realff, and Yoshiaki Kawajiri.
\newblock Systems {Design} and {Economic} {Analysis} of {Direct} {Air} {Capture} of {CO} $_{\textrm{2}}$ through {Temperature} {Vacuum} {Swing} {Adsorption} {Using} {MIL}-101({Cr})-{PEI}-800 and mmen-{Mg} $_{\textrm{2}}$ (dobpdc) {MOF} {Adsorbents}.
\newblock {\em Industrial \& Engineering Chemistry Research}, 56(3):750--764, January 2017.

\bibitem{leonzio_environmental_2022}
Grazia Leonzio, Onesmus Mwabonje, Paul~S. Fennell, and Nilay Shah.
\newblock Environmental performance of different sorbents used for direct air capture.
\newblock {\em Sustainable Production and Consumption}, 32:101--111, July 2022.

\bibitem{wurzbacher_heat_2016}
Jan~Andre Wurzbacher, Christoph Gebald, Samuel Brunner, and Aldo Steinfeld.
\newblock Heat and mass transfer of temperature–vacuum swing desorption for {CO2} capture from air.
\newblock {\em Chemical Engineering Journal}, 283:1329--1338, January 2016.

\bibitem{marinic_direct_2023}
Dana Marinič and Blaž Likozar.
\newblock Direct air capture multiscale modelling: {From} capture material optimization to process simulations.
\newblock {\em Journal of Cleaner Production}, 408:137185, July 2023.

\bibitem{keith_process_2018}
David~W. Keith, Geoffrey Holmes, David St.~Angelo, and Kenton Heidel.
\newblock A {Process} for {Capturing} {CO2} from the {Atmosphere}.
\newblock {\em Joule}, 2(8):1573--1594, August 2018.

\bibitem{schafer_towards_2024}
Malte Schäfer, Felipe Cerdas, and Christoph Herrmann.
\newblock Towards standardized grid emission factors: methodological insights and best practices.
\newblock {\em Energy \& Environmental Science}, 17(8):2776--2786, 2024.

\bibitem{ruggles_developing_2020}
Tyler~H. Ruggles, David~J. Farnham, Dan Tong, and Ken Caldeira.
\newblock Developing reliable hourly electricity demand data through screening and imputation.
\newblock {\em Scientific Data}, 7(1):155, May 2020.

\bibitem{ruhnau_flexible_2023}
Oliver Ruhnau and Johanna Schiele.
\newblock Flexible green hydrogen: {The} effect of relaxing simultaneity requirements on project design, economics, and power sector emissions.
\newblock {\em Energy Policy}, 182:113763, November 2023.

\bibitem{bachman_rail-based_2022}
E.~Bachman, Alexandra Tavasoli, T.~Alan Hatton, Christos~T. Maravelias, Erik Haites, Peter Styring, Alán Aspuru-Guzik, Jeffrey MacIntosh, and Geoffrey Ozin.
\newblock Rail-based direct air carbon capture.
\newblock {\em Joule}, 6(7):1368--1381, July 2022.

\bibitem{groom_social_2023}
Ben Groom and Frank Venmans.
\newblock The social value of offsets.
\newblock {\em Nature}, 619(7971):768--773, July 2023.

\bibitem{shah_cement_2022}
Izhar~Hussain Shah, Sabbie~A. Miller, Daqian Jiang, and Rupert~J. Myers.
\newblock Cement substitution with secondary materials can reduce annual global {CO2} emissions by up to 1.3 gigatons.
\newblock {\em Nature Communications}, 13(1):5758, September 2022.

\bibitem{bhardwaj_opportunities_2021}
Amar Bhardwaj, Colin McCormick, and Julio Friedmann.
\newblock Opportunities and {Limits} of {CO2} {Recycling} in a {Circular} {Carbon} {Economy}: {Techno}-economics, {Critical} {Infrastructure} {Needs}, and {Policy} {Priorities}, May 2021.

\bibitem{us_congress_inflation_2022}
U.S. Congress.
\newblock Inflation {Reduction} {Act} of 2022, 2022.

\bibitem{sun_impact_2022}
Liang Sun and Wenying Chen.
\newblock Impact of carbon tax on {CCUS} source-sink matching: {Finding} from the improved {ChinaCCS} {DSS}.
\newblock {\em Journal of Cleaner Production}, 333:130027, January 2022.

\bibitem{victor_ccus_2022}
Nadejda Victor and Christopher Nichols.
\newblock {CCUS} deployment under the {U}.{S}. {45Q} tax credit and adaptation by other {North} {American} {Governments}: {MARKAL} modeling results.
\newblock {\em Computers \& Industrial Engineering}, 169:108269, July 2022.

\bibitem{low_analytical_2023}
May-Yin~(Ashlyn) Low, Lucy~Victoria Barton, Ronny Pini, and Camille Petit.
\newblock Analytical review of the current state of knowledge of adsorption materials and processes for direct air capture.
\newblock {\em Chemical Engineering Research and Design}, 189:745--767, January 2023.

\bibitem{lackner_buying_2021}
Klaus~S. Lackner and Habib Azarabadi.
\newblock Buying down the {Cost} of {Direct} {Air} {Capture}.
\newblock {\em Industrial \& Engineering Chemistry Research}, 60(22):8196--8208, June 2021.

\bibitem{sukprasert_implications_2024}
Thanathorn Sukprasert, Noman Bashir, Abel Souza, David Irwin, and Prashant Shenoy.
\newblock On the {Implications} of {Choosing} {Average} versus {Marginal} {Carbon} {Intensity} {Signals} on {Carbon}-aware {Optimizations}.
\newblock In {\em The 15th {ACM} {International} {Conference} on {Future} and {Sustainable} {Energy} {Systems}}, pages 422--427, Singapore Singapore, June 2024. ACM.

\bibitem{copernicus_climate_change_service_era5_2018}
{Copernicus Climate Change Service}.
\newblock {ERA5} hourly data on pressure levels from 1940 to present, 2018.

\end{thebibliography}



\end{document}